\documentclass[reprint,superscriptaddress,amsmath,amssymb,aps,prl]{revtex4-2}

\usepackage[T1]{fontenc}
\usepackage[utf8]{inputenc}
\usepackage{lmodern}
\usepackage{microtype}
\usepackage{needspace}

\usepackage{amsmath,amssymb,bm}
\usepackage{graphicx}
\graphicspath{{figures/}{../figures/}}
\usepackage{physics}
\usepackage{xcolor}
\usepackage{hyperref}
\usepackage{orcidlink}
\hypersetup{hidelinks}

\begin{document}

\title{Collective Strong Quantum Mpemba Effect in Accelerated Quantum Detectors}

\author{Richard Q. Matos}
\affiliation{
Instituto de F\'isica,
Universidade Federal de Goi\'as,
74690-900, Goi\^ania, Goi\'as, Brazil
}

\author{G. D. de Moraes Neto~\orcidlink{0000-0003-4273-8380}}
\email{gentil.neto@hibiuh.edu.cn}
\affiliation{Department of Fundamental Sciences, Hainan Bielefeld University of Applied Sciences, Danzhou, Hainan 578101, China}

\author{Norton G. de Almeida~\orcidlink{0000-0001-8517-6774}}
\email{norton@ufg.br}
\affiliation{Instituto de F\'isica, Universidade Federal de Goi\'as, 74690-900, Goi\^ania--GO, Brazil}

\date{\today}

\begin{abstract}
	Collective coupling to the Minkowski vacuum allows uniformly accelerated
	quantum detectors to exhibit a strong quantum Mpemba effect, in which a
	state initially farther from equilibrium relaxes faster than a closer
	one. We establish the existence of this effect for every fixed finite
	number of detectors $N\geq 2$ at sufficiently small transverse
	separation within the pointlike secular master equation. The mechanism
	originates from two reflection sectors whose populations are conserved
	by the leading collective dynamics. Weak intersector transfer at finite
	separation generates an isolated, parametrically slow relaxation mode.
	A simple family of initial states permits its leading-order cancellation
	by matching the equilibrium sector populations, while a corrected
	preparation eliminates the slow-mode amplitude exactly. We construct
	a reference state that establishes a genuine reversal of trace-distance
	ordering. Numerical calculations and local-rate-preserving controls
	identify the collective origin of the effect, while spatially smeared
	detector calculations support its persistence within the second-order
	secular framework. The same mechanism extends to detectors coupled to
	a common thermal field, with analytically distinct slow relaxation rates.
	Thus, strong Mpemba relaxation is not exclusive to acceleration but
	emerges from a more general interplay between collective dissipation,
	reflection-sector conservation, and weak symmetry-breaking transfer.
\end{abstract}

\maketitle

\emph{Introduction.---}
The Mpemba effect describes the counterintuitive situation in which
a state initially farther from equilibrium relaxes faster than a
closer one \cite{mpemba1969cool,summer2026resource}.
Its strong form arises when an initial preparation eliminates the
slowest relaxation mode, leading to a faster asymptotic approach
to stationarity \cite{KlichEtAl2019,CarolloLasantaLesanovsky2021}.
Quantum analogues have attracted considerable attention, including
an experimental demonstration of the strong effect
\cite{ZhangEtAlExperiment2025}.
In isolated quantum many-body systems, related phenomena concern
the relaxation of subsystem observables or the restoration of
symmetries rather than convergence of the full density operator
to equilibrium \cite{ares2025quantum,yu2025quantum}.

In open quantum systems, different relaxation mechanisms and
preparation strategies have been explored
\cite{caceffo2024entangled,dong2025quantum,
	ZhangLuoWu2026,bagui2026accelerated}.
In particular, weak symmetries can be used to prepare states
that exclude symmetry-odd slow modes
\cite{westhoff2025fast}, while weakly coupled sectors can
produce metastable relaxation that suitable preparations
bypass \cite{longhi2026giant}.
Strong Mpemba relaxation has also been investigated in
long-range spin chains subject to dephasing
\cite{WeiEtAl2026,SongEtAl2026}.
These developments motivate identifying how a microscopic
environment generates the slow relaxation modes and determines
which initial states can avoid them.

Uniformly accelerated detectors provide a setting in which
the environment is the Minkowski vacuum itself.
A detector with proper acceleration $a$ responds thermally
at the Unruh temperature
$T_U=\hbar a/(2\pi c k_B)$ \cite{unruh1976notes}.
Its thermal response, however, does not fully characterize
the relaxation dynamics, which depends on the field
correlations sampled along its trajectory.
Mpemba-like thermalization has recently been reported for
a single accelerated detector \cite{WangEtAlUnruh2026}.
For multiple detectors, correlations between distinct
trajectories generate collective dissipation and
field-mediated interactions
\cite{richter2017collective,KaplanekTjoa2023},
and can enhance cooperative responses such as the
superradiant Unruh signal \cite{DeswalEtAl2025}.
Whether these correlations also produce an isolated slow
relaxation mode that can be eliminated by a physical
initial preparation remains a distinct question.

Here we establish a collective strong quantum Mpemba effect
for every fixed finite number of detectors $N\geq2$,
at sufficiently small transverse separation within the
pointlike secular master equation.
The field-mediated exchange and collective dissipation
leave precisely two invariant reflection sectors in the
leading dynamics.
At finite separation, weak intersector transfer produces
an isolated slow mode associated with the redistribution
of their populations.
Matching the equilibrium reflection-sector weights
eliminates this mode to leading order, and a corrected
preparation cancels its amplitude exactly.
Together with an explicitly constructed reference state,
this establishes a genuine reversal of trace-distance
ordering.
A four-detector example, local-rate-preserving controls,
and spatially smeared calculations illustrate the
mechanism and its scope.
Comparison with a common thermal field further reveals
how different spatial correlations modify the collective
relaxation rates despite identical equilibrium states.
Unlike weak-symmetry selection of reflection-odd modes,
the present mechanism eliminates a slow population
imbalance between reflection sectors.
\begin{figure*}[t]
	\centering
	\includegraphics[width=0.99\textwidth]{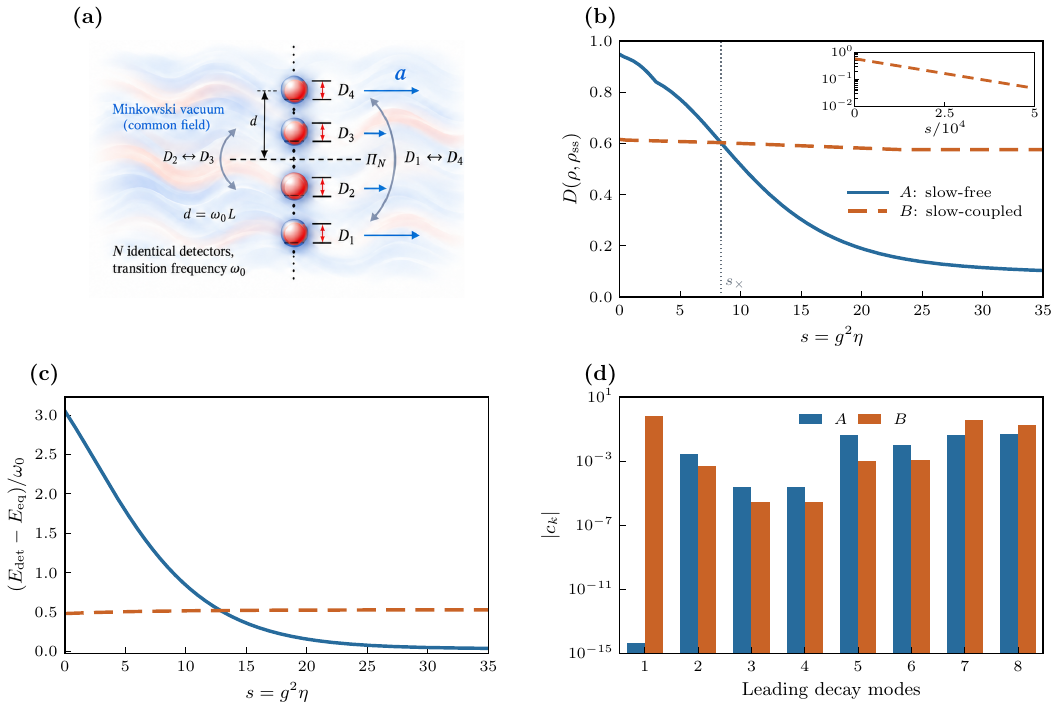}
	\caption{\textbf{Collective strong quantum Mpemba relaxation in an
			accelerated detector chain.} (a) Four identical detectors coupled to
		the Minkowski vacuum, with acceleration transverse to their array.
		Reflection $\Pi_4$ exchanges $D_1\leftrightarrow D_4$ and
		$D_2\leftrightarrow D_3$; adjacent sites are separated by
		$d=\omega_0L$ in dimensionless transverse coordinates.
		(b)--(d) Pointlike secular dynamics for $N=4$, $E=\omega_0/a=0.424264$,
		$d=0.01$, and $s=g^2\eta$. (b) Trace distance for optimized
		preparation $A$ (solid blue) and reference $B$ (dashed orange).
		Despite $D_A(0)>D_B(0)$, the curves cross at $s_\times\simeq8.3674$.
		The inset follows $B$ to $s=5\times10^4$, revealing finite slow
		relaxation, $\bar\gamma_{\rm slow}\simeq5.06156\times10^{-5}$.
		(c) Excess detector energy $(E_{\rm det}-E_{\rm eq})/\omega_0$.
		(d) Modal magnitudes in the eight slowest nonstationary modes of the
		zero-excitation-difference operator block, with $\|R_k\|_{\rm HS}=1$.
		$A$ eliminates the slowest component to numerical precision, while
		$|c_1^{(B)}|\simeq0.631$; values below $10^{-15}$ are displayed at
		this floor.}
	\label{fig:mpemba_main}
	\label{fig:mechanism}
\end{figure*}


\emph{Collective accelerated-detector model.—}
We consider $N$ identical two-level Unruh--DeWitt detectors with transition
frequency $\omega_0$, collectively coupled to a common massless scalar field in the Minkowski vacuum.
Henceforth we use natural units $\hbar=c=k_B=1$.
The free detector Hamiltonian is
\begin{equation}
    H_S
    =
    \frac{\omega_0}{2}
    \sum_{i=1}^{N}
    \sigma_i^z .
    \label{eq:HS}
\end{equation}

The detectors accelerate along the $x$ direction with the same proper
acceleration $a$ and are arranged in a straight line along the transverse
$y$ direction. Their worldlines, in Minkowski coordinates $(t,x,y,z)$, are
\begin{equation}
\begin{aligned}
    x_i^\mu(\tau)
    &=\left(
        \frac{\sinh(a\tau)}{a},
        \frac{\cosh(a\tau)}{a},
        (i-1)L,0
    \right),\\
    &\qquad i=1,\ldots,N .
\end{aligned}
    \label{eq:worldlines}
\end{equation}
Here $L$ is the proper separation between neighboring detectors; detectors
$i$ and $j$ are separated by $|i-j|L$. All detectors therefore share the
same proper time $\tau$ and remain at fixed relative positions in the
accelerated frame while sampling distinct, correlated regions of the
Minkowski vacuum.

It is convenient to express the dynamics in the common Rindler time $ \eta=a\tau .$
The corresponding free generator is
$H_\eta=(E/2)\sum_{i=1}^N\sigma_i^z$, where $E\equiv\omega_0/a$.

The detector--field interaction in Rindler time is
\begin{equation}
    H_I(\eta)
    =
    g
    \sum_{i=1}^{N}
    \xi\,
    \mu_i(\eta)\,
    \phi[x_i(\eta)],
    \qquad
    \xi=\frac{1}{a},
    \label{eq:HI}
\end{equation}
where $g\ll1$ is the dimensionless detector--field coupling,
$\phi$ is the massless scalar field, and
\begin{equation}
    \mu_i(\eta)
    =
    \sigma_i^+ e^{iE\eta}
    +
    \sigma_i^- e^{-iE\eta}
\end{equation}
is the transition monopole operator of detector $i$ in the interaction
picture. The worldline $x_i(\eta)$ is
given by Eq.~\eqref{eq:worldlines} at $\tau=\eta/a$. In these units,
$T_U=a/(2\pi)$ and $\beta_U=2\pi/a$,
hence $\beta_U\omega_0=2\pi E$. Within the Born--Markov and
secular approximations, the reduced dynamics is described by
\begin{align}
    \frac{d\rho}{d\eta}
    ={}&
    -i[H_\eta+H_{\rm LS},\rho]
    \nonumber\\
    &+
    \sum_{i,j}
    \Gamma_{ij}^{\downarrow}
    \left(
        \sigma_j^-\rho\sigma_i^+
        -
        \frac12
        \left\{
            \sigma_i^+\sigma_j^-,
            \rho
        \right\}
    \right)
    \nonumber\\
    &+
    \sum_{i,j}
    \Gamma_{ij}^{\uparrow}
    \left(
        \sigma_j^+\rho\sigma_i^-
        -
        \frac12
        \left\{
            \sigma_i^-\sigma_j^+,
            \rho
        \right\}
    \right).
    \label{eq:GKLS}
\end{align}
Here $H_{\rm LS}$ is the field-induced Lamb-shift Hamiltonian,
which includes local frequency shifts and coherent interactions
between detectors. The diagonal rates
$\Gamma_{ii}^{\uparrow,\downarrow}$ describe local excitation
and decay and satisfy the Unruh detailed-balance relation
$\Gamma_{ii}^{\uparrow}/\Gamma_{ii}^{\downarrow}=e^{-\beta_U\omega_0}=e^{-2\pi E}$,
whereas the off-diagonal coefficients
$\Gamma_{ij}^{\uparrow,\downarrow}$ with $i\neq j$ encode
field-mediated collective dissipation. In the convention used here,
\begin{equation}
    \Gamma_{ij}^{\downarrow}
    =g^2\xi^2\mathcal G_{ij}(E),
    \qquad
    \Gamma_{ij}^{\uparrow}
    =g^2\xi^2\mathcal G_{ij}(-E),
    \label{eq:rates_spectra}
\end{equation}
where the field-correlation spectrum in common Rindler time is
\begin{equation}
    \mathcal G_{ij}(\omega)
    =
    \int_{-\infty}^{\infty}
    d\Delta\eta\,
    e^{i\omega\Delta\eta}
    G^+_{ij}(\Delta\eta),
    \label{eq:Gij}
\end{equation}
with
\begin{equation}
    G^+_{ij}(\Delta\eta)
    =
    \langle 0_M|
    \phi[x_i(\eta)]\phi[x_j(\eta')]
    |0_M\rangle,
    \qquad
    \Delta\eta=\eta-\eta'.
    \label{eq:Wightman}
\end{equation}
The derivation of
Eq.~\eqref{eq:GKLS}, including these factors, the coherent couplings,
and the approximations used, is given in the Supplemental Material \cite{SupplementalMaterial}.

For the parameter regimes considered below, the stationary state is
\begin{equation}
    \rho_{\rm ss}
    =
    \frac{
        e^{-\beta_U H_S}
    }{
        \Tr(e^{-\beta_U H_S})
    }
    =
    \frac{
        e^{-2\pi H_\eta}
    }{
        \Tr(e^{-2\pi H_\eta})
    }.
    \label{eq:ss}
\end{equation}

We denote the Rindler-time generator by $d\rho/d\eta=\mathcal L_a[\rho]$.
Acceleration controls both the Unruh thermal factors and
the collective dissipative structure. Figure~\ref{fig:mechanism}(a) illustrates the geometry and reflection symmetry;
its remaining panels display the representative numerical dynamics.

\emph{Reflection-controlled slow relaxation.---}
Let $\Pi_N$ reverse the detector order and
$P_\pm=(\mathbb I\pm\Pi_N)/2$. We use $d=\omega_0L$ and the reduced
time $s=g^2\eta$. At fixed finite $N$ and $E>0$, the leading coherent
exchange together with $J_\pm=\sum_i\sigma_i^\pm$ has commutant
$\operatorname{span}\{\mathbb I,\Pi_N\}$, as proved in the Supplemental
Material \cite{SupplementalMaterial}. The collective reference dynamics therefore preserves two
reflection sectors \cite{buvca2012note}. Small spatial corrections weakly mix them, producing
a simple real slow eigenvalue with reduced decay rate
\begin{equation}
 \bar\gamma_{\rm slow}=\kappa_{N,E}d^2+O(d^4),
 \qquad \kappa_{N,E}>0.
 \label{eq:fn_slow_rate}
\end{equation}
All other nonstationary modes retain decay rates bounded away
from zero as $d\to0$ at fixed $N$ and $E$, yielding the spectral
separation underlying metastable relaxation
\cite{macieszczak2016towards}. The slow mode describes the redistribution between
sector states $\rho_\pm=P_\pm\rho_{\rm ss}/w_\pm$, where
$w_\pm=(1\pm t_N)/2$ and
\begin{equation}
 t_N=\operatorname{Tr}(\Pi_N\rho_{\rm ss})
 =\left[\frac{1+q^2}{(1+q)^2}\right]^{\lfloor N/2\rfloor},
 \qquad q=e^{-2\pi E}.
 \label{eq:fn_thermal_parity}
\end{equation}
The slow left eigenoperator satisfies
$\ell_{\rm slow}=(\Pi_N-t_N I)/2+O(d^2)$.
Here reflection is a strong symmetry of the leading dynamics: its
Hilbert-space sector populations are conserved. At finite separation,
the generator still commutes with reflection conjugation as a superoperator,
but reflection-odd jump channels transfer population between those sectors.
The slow observable is itself reflection-even, in contrast to the
reflection-odd modes excluded by weak-symmetry state selection
\cite{westhoff2025fast}. Its cancellation requires adjusting the sector
weights, rather than preparing a reflection-invariant density matrix.
Thus, for an arbitrary initial state, the leading cancellation
condition is $\operatorname{Tr}(\Pi_N\rho)=t_N$;
exact cancellation at finite separation requires matching the
corrected slow observable.
For $\rho_A^{(N)}(p)=p|1^N\rangle\langle1^N|
+(1-p)|10^{N-1}\rangle\langle10^{N-1}|$, the initial reflection
expectation is $p$. Matching the equilibrium sector weights thus gives
the leading cancellation rule. More precisely, a unique physical value
\begin{equation}
 p_\star(d)=t_N+O(d^2),\qquad
 \mathcal P_{\rm slow}[\rho_A^{(N)}(p_\star)-\rho_{\rm ss}]=0
 \label{eq:fn_exact_cancellation}
\end{equation}
exists for sufficiently small $d$. Here
$\mathcal P_{\rm slow}(X)=R_{\rm slow}
\operatorname{Tr}(L_{\rm slow}^{\dagger}X)$, with left and right slow
eigenoperators normalized by
$\operatorname{Tr}(L_{\rm slow}^{\dagger}R_{\rm slow})=1$.
An explicit reference state, constructed in the Supplemental Material \cite{SupplementalMaterial},
starts closer to equilibrium but retains this slow component, proving a
trace-distance crossing. The allowed separation depends on $N,E$; no
interval uniform in $N$ is asserted. We next illustrate the mechanism
using a separately chosen numerical reference state.

\begin{figure*}[t]
    \centering
    \includegraphics[
        width=\textwidth
    ]{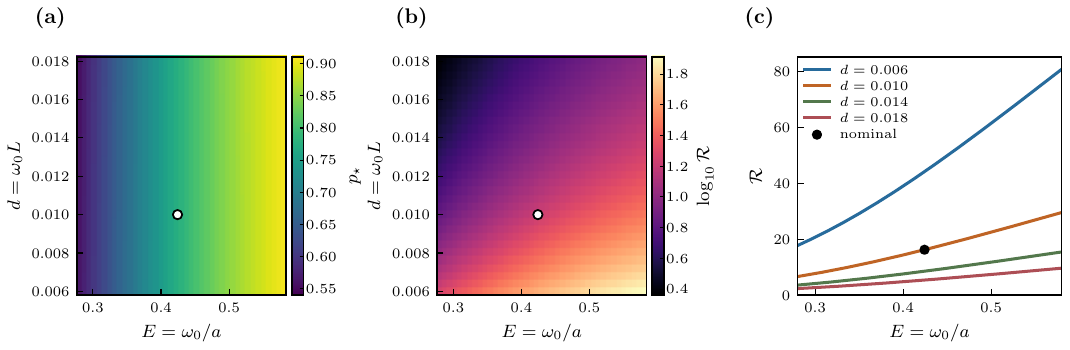}
    \caption{
    \textbf{Acceleration and geometry dependence of the strong-Mpemba
    mechanism.}
    (a) Strong-Mpemba preparation parameter $p_\star$ as a function of
    $E=\omega_0/a$ and the transverse detector separation $d$.
    (b) Corresponding spectral enhancement
    $\log_{10}\mathcal R$, where
    $\mathcal R=\gamma_{k_*}/\gamma_{\rm slow}$.
    (c) Cuts of $\mathcal R$ as a function of $E$ for different detector
    separations. The pointlike secular grid comprises $41\times31$ directly evaluated
    points for $0.28\leq E\leq0.58$ and $0.006\leq d\leq0.018$. The marker denotes the reference (nominal)
    operating point $E=0.424264$ and $d=0.01$, for which
    $p_\star\simeq0.771543$ and $\mathcal R_4\simeq16.25$.
    }
    \label{fig:acceleration_dependence}
\end{figure*}


\emph{Collective strong Mpemba relaxation under acceleration.—}
We now specialize to $N=4$ at the representative operating point
$E=0.424264$ and $d=0.01$, where $d=\omega_0L$ denotes
the dimensionless nearest-neighbor transverse separation.

\Needspace{3\baselineskip}
We use this point as a representative example within the sampled
parameter region: it admits a physical slow-free preparation and
a well-resolved trace-distance crossing, while the broader
dependence on $E$ and $d$ is shown in
Fig.~\ref{fig:acceleration_dependence}.

\Needspace{8\baselineskip}
The strong-Mpemba state is prepared within the one-parameter family
\begin{equation}
    \rho_A(p)
    =
    p\,|1111\rangle\langle1111|
    +
    (1-p)\,|1000\rangle\langle1000|.
    \label{eq:rhoA_family}
\end{equation}
Projecting this family onto the slow Liouvillian band determines
$p_\star=0.7715429438$. Its slow-band projection vanishes to
numerical precision.

\Needspace{12\baselineskip}
For comparison, we choose a state that starts closer to equilibrium but
retains a finite slow-mode component,
\begin{equation}
    \rho_B
    =
    (1-r)\rho_{\rm ss}
    +
    r\,|F_2\rangle\langle F_2|,
    \label{eq:rhoB}
\end{equation}
with
\begin{equation}
    |F_2\rangle
    =
    \frac{1}{2}
    \left(
        |1000\rangle
        -
        |0100\rangle
        +
        |0010\rangle
        -
        |0001\rangle
    \right),
    \label{eq:F2}
\end{equation}
and $r=0.6499877261$.
The initial trace distances are $D_A(0)=0.946834$ and
$D_B(0)=0.615442$, where
\begin{equation}
    D_X(s)
    =
    \frac12
    \left\|
        \rho_X(s)-\rho_{\rm ss}
    \right\|_1.
    \label{eq:trace_distance}
\end{equation}
\Needspace{10\baselineskip}
Thus state $A$ begins substantially farther from equilibrium. Nevertheless, in terms of the rescaled Rindler time
$s=g^2\eta$, the ordering reverses at $s_\times \simeq 8.3674,$
after which $D_A(s)<D_B(s)$.

Figure~\ref{fig:mpemba_main} shows this reversal in the trace distance
[panel (b)] and the relaxation of the detector energy for the same
initial states [panel (c)]. The modal projections in panel (d)
identify its spectral origin: state $A$ nearly eliminates the
slowest decaying contribution, whereas state $B$ retains it.

The crossing has a direct spectral origin. Labeling $k_*$
as the slowest remaining decay mode with nonzero weight
in the ideal optimized state, at the reference point the
reduced decay rates $\bar{\gamma}_k\equiv\gamma_k/g^2$ are
\begin{equation}
    \bar{\gamma}_{\rm slow}
    =
    5.06156\times10^{-5},
    \qquad
    \bar{\gamma}_{k_*}
    =
    8.22472\times10^{-4}.
    \label{eq:nominal_rates}
\end{equation}
The nearly horizontal reference trajectory in Fig.~\ref{fig:mpemba_main}(b)
thus does not represent a dark state. Its slow relaxation time,
$1/\bar\gamma_{\rm slow}\simeq1.98\times10^4$, greatly exceeds the
plotted interval $0\leq s\leq35$; the displayed trajectory need not
have reached its asymptotic single-mode regime.
For the numerical modal amplitudes, we use
biorthogonal eigenoperators satisfying
$\operatorname{Tr}(L_k^\dagger R_{k'})=\delta_{kk'}$
and fix the right eigenoperators by
$\|R_k\|_{\rm HS}=1$.
The coefficients are
$c_k^{(X)}=\operatorname{Tr}[L_k^\dagger(\rho_X-\rho_{\rm ss})]$,
where $X=A,B$ labels the initial state.
The corresponding slow-mode amplitudes satisfy
\begin{equation}
    |c_{\rm slow}^{(A)}|
    \lesssim
    10^{-10},
    \qquad
    |c_{\rm slow}^{(B)}|
    \simeq
    6.31\times10^{-1}.
    \label{eq:slow_overlaps}
\end{equation}
State $A$ therefore bypasses the slowest collective channel, whereas state
$B$ retains it.
In the ideal slow-free preparation, the leading asymptotic terms are
\begin{align}
    \rho_B(\eta)-\rho_{\rm ss}
    &\sim
    c_{\rm slow}^{(B)}
    e^{-\gamma_{\rm slow}\eta}
    R_{\rm slow},
    \nonumber\\
    \rho_A(\eta)-\rho_{\rm ss}
    &\sim
    c_{k_*}^{(A)}
    e^{-\gamma_{k_*}\eta}
    R_{k_*}.
    \label{eq:asymptotic_modes}
\end{align}
The resulting enhancement of the asymptotic relaxation rate is
\begin{equation}
    \mathcal R_4
    \equiv
    \frac{
        \gamma_{k_*}
    }{
        \gamma_{\rm slow}
    }
    \simeq
    16.25.
    \label{eq:R4}
\end{equation}


Acceleration and detector geometry control both the Liouvillian spectrum
and the orientation of its slow eigenspaces relative to the accessible
initial states. To characterize this dependence, we determine
$p_\star(E,d)$ and the spectral enhancement
$\mathcal R(E,d)=\gamma_{k_*}(E,d)/\gamma_{\rm slow}(E,d)$.
The former identifies the physical state within the family
\eqref{eq:rhoA_family} that suppresses the slow band, while the latter
quantifies the resulting separation between the effective strong-Mpemba
decay scale and the slowest generic relaxation scale.



\emph{Discussion and outlook.---}
The collective relaxation mechanism is governed by the population
imbalance between reflection sectors. At finite separation, this
imbalance is not conserved but relaxes through a weak intersector
channel. Preparing the equilibrium sector populations suppresses
the associated slow mode to leading order, while a finite-separation
correction eliminates its amplitude exactly. The resulting strong
Mpemba effect is therefore a reordering of relaxation times rather
than an absolute speedup over independent detectors. This distinction
is supported by a control model in which the cross rates and coherent
exchange are removed while the local rates are preserved: the chosen
preparation family no longer cancels the collective slow mode, and
the reference pair exhibits no crossing over $0\leq s\leq400$.

The parameter maps in Fig.~\ref{fig:acceleration_dependence}
show that the optimal preparation depends primarily on the
acceleration parameter $E$, with a weaker dependence on separation,
as expected from Eq.~\eqref{eq:fn_thermal_parity}. The relaxation-rate
enhancement increases as the detectors approach the collective limit
within the sampled parameter range, consistently with the
small-separation prediction. Spatially smeared detector calculations
also retain slow-mode suppression and trace-distance crossing within
the second-order secular framework
\cite{SupplementalMaterial}. These results support the persistence
of the mechanism beyond ideal pointlike detectors, although they do
not establish bounds on microscopic finite-coupling corrections
relative to the small collective decay rate.

The mechanism is not exclusive to accelerated detectors. An inertial
common thermal field at the Unruh temperature $T_U$ admits the same
finite-$N$ construction and leading-order preparation rule. Although
the stationary state and equilibrium reflection populations coincide,
the spatial field correlations differ, producing distinct rates of
slow intersector relaxation. In the pointlike small-separation limit,
we obtain
\begin{equation}
	\lim_{d\to0^+}
	\frac{\bar\gamma_{\rm slow}^{U}(d)}
	{\bar\gamma_{\rm slow}^{\rm th}(d)}
	=1+E^{-2}
	=1+\frac{a^2}{\omega_0^2}.
	\label{eq:slow_rate_environment_ratio}
\end{equation}
For each fixed finite $N$, this ratio is independent of detector
number and compares the slow relaxation rates, rather than the
strong-Mpemba enhancement factors. Acceleration thus modifies the
collective relaxation dynamics even when local detailed balance
and the stationary equilibrium state are identical.

More generally, the separation between rapid intrasection
equilibration and slow reflection-sector transfer provides a route
to strong Mpemba relaxation in correlated quantum environments
without invoking acceleration. The present results establish its
accelerated-detector realization and identify how the microscopic
field correlations determine the slow relaxation rate. Extending
this picture to additional local noise and imperfect reflection
symmetry, which modify the slow mode and its preparation condition,
remains an important direction for further investigation.

\begin{acknowledgments}
The authors acknowledge financial support from CAPES,
CNPq, and FAPESP, Brazilian funding agencies.

The authors used OpenAI ChatGPT to assist in developing
and checking the analytical arguments presented in the
Supplemental Material~\cite{SupplementalMaterial}, and
in drafting and revising scientific explanations.
All resulting text was reviewed and edited by the authors,
who take full responsibility for the correctness and
originality of the results and for the final manuscript.
\end{acknowledgments}

\par\noindent\textit{Data availability.} The numerical data supporting the results and the custom scripts used for the reported calculations are available from the corresponding author upon reasonable request.

\IfFileExists{references.bib}{\bibliography{references}}{\bibliography{manuscript/references}}

\end{document}


\begin{center}
{\large\bfseries Supplemental Material for\\[3pt]
``Collective Strong Quantum Mpemba Effect in Accelerated Quantum Detectors''}
\par\vspace{8pt}
Richard Q. Matos$^{1}$, G. D. de Moraes Neto$^{2}$, and Norton G. de Almeida$^{1}$
\par\vspace{5pt}
{\small $^{1}$Instituto de F\'isica, Universidade Federal de Goi\'as,
74690-900, Goi\^ania, Goi\'as, Brazil\\
$^{2}$Department of Fundamental Sciences, Hainan Bielefeld University of
Applied Sciences, Danzhou, Hainan 578101, China}
\end{center}
\tableofcontents
\bigskip
\noindent This Supplemental Material specifies the generator and numerical
construction used in the Letter, presents the local and thermal controls,
states the scope of the spatial-profile checks, and proves the finite-$N$
reflection and strong-Mpemba results. Natural units $\hbar=c=k_B=1$ are used.

\section{Microscopic generator and stationary state}
\label{sec:model}
\subsection{Transverse trajectories and field response}
We consider $N$ identical Unruh--DeWitt detectors of proper gap $\omega_0$,
with common acceleration $a$ and transverse nearest-neighbor separation $L$.
The common Rindler time is $\eta=a\tau$, and
\begin{equation}
 x_i^\mu(\eta)=\left(a^{-1}\sinh\eta,a^{-1}\cosh\eta,(i-1)L,0\right),
 \qquad E=\frac{\omega_0}{a},\qquad d=\omega_0L.
 \label{eq:geometry}
\end{equation}
With $\sigma^-=|0\rangle\langle1|$ and $\sigma^z=[\sigma^+,\sigma^-]$,
the free and interaction Hamiltonians in Rindler time are
\begin{equation}
 H_\eta=\frac{E}{2}\sum_i\sigma_i^z,\qquad
 H_I(\eta)=\frac{g}{a}\sum_i
 \left(\sigma_i^+e^{iE\eta}+\sigma_i^-e^{-iE\eta}\right)\phi[x_i(\eta)].
 \label{eq:interaction}
\end{equation}
The weak-coupling construction follows the standard detector and
open-system treatments~\cite{unruh1976notes,benatti2004entanglement,crispino2008unruh,breuer2002theory}.
For the massless scalar field in the Minkowski vacuum, the pulled-back
Wightman function is
\begin{equation}
 G^+_{ij}(v)=-\frac{a^2}{16\pi^2}
 \frac{1}{\sinh^2[(v-i\epsilon)/2]-\sinh^2(\chi_{ij}/2)},
 \qquad
 \chi_{ij}=2\operatorname{arsinh}\frac{a|i-j|L}{2}
 =2\operatorname{arsinh}\frac{|i-j|d}{2E}.
 \label{eq:wightman}
\end{equation}
Here $v=\eta-\eta'$. The stationary $i\epsilon$ prescription is understood
as the pointlike limit of a spatially regularized detector; it yields the
standard stationary rates~\cite{schlicht2004considerations,louko2006often}.
Born--Markov and secular reduction retains all resonant cross terms because
the detectors have equal Rindler gaps~\cite{breuer2002theory,lindblad1976generators,gorini1976completely}.
Contour integration of
$\Gamma^\downarrow_{ij}=g^2a^{-2}\int_{\mathbb R}dv\,e^{iEv}G^+_{ij}(v)$
and its negative-frequency counterpart gives
\begin{equation}
 \frac{\Gamma^\downarrow_{ij}}{g^2}=a_\downarrow F_{ij},\qquad
 \frac{\Gamma^\uparrow_{ij}}{g^2}=a_\uparrow F_{ij},\qquad
 a_\downarrow=\frac{E}{2\pi(1-q)},\quad a_\uparrow=qa_\downarrow,\quad
 q=e^{-2\pi E},\qquad
 F_{ij}=\frac{\sin(E\chi_{ij})}{E\sinh\chi_{ij}},\quad F_{ii}=1.
 \label{eq:rates}
\end{equation}
Thus the complete matrices obey $\Gamma^\uparrow=q\Gamma^\downarrow$,
with $q=e^{-\beta_U\omega_0}$ and $\beta_U=2\pi/a$.
The Gram construction in Sec.~\ref{fn:dyn:app:gram} proves $F\geq0$ directly.

\subsection{Coherent exchange and reduced generator}
The Lamb-shift Hamiltonian contains local shifts and coherent exchange,
\begin{equation}
 H_{\rm LS}=\sum_i\frac{\delta\omega_i}{2}\sigma_i^z
       +\sum_{i\ne j}J^U_{ij}\sigma_i^+\sigma_j^-,\qquad
 J^U_{ij}=-\frac{g^2\cos(E\chi_{ij})}{4\pi\sinh\chi_{ij}}.
 \label{eq:exchange}
\end{equation}
Both resonant Bohr frequencies contribute to the hopping coefficient.
Explicitly, with $\widetilde G_{ij}=a^{-2}G^+_{ij}$ and
$S_{ij}(\omega)=\operatorname{Im}\int_0^\infty dv\,e^{i\omega v}
\widetilde G_{ij}(v)$,
\begin{equation}
 J^U_{ij}=g^2[S_{ij}(E)+S_{ij}(-E)],\qquad
 \widetilde{\mathcal G}_{ij}(\nu)-\widetilde{\mathcal G}_{ij}(-\nu)
 =\frac{\sin(\nu\chi_{ij})}{2\pi\sinh\chi_{ij}}.
\end{equation}
The symmetric thermal contribution cancels in this sum of principal-value
transforms. A common local shift is absorbed into the renormalized gap.

In the interaction picture and reduced time $s=g^2\eta$, write
$d\rho/ds=\mathcal L_d\rho$, where
\begin{align}
 \mathcal L_d\rho={}&-i[\bar H(d),\rho]
 +\sum_{ij}F_{ij}\left[
 a_\downarrow\left(\sigma_j^-\rho\sigma_i^+
 -\tfrac12\{\sigma_i^+\sigma_j^-,\rho\}\right)
 +a_\uparrow\left(\sigma_j^+\rho\sigma_i^-
 -\tfrac12\{\sigma_i^-\sigma_j^+,\rho\}\right)\right],
 \label{fn:dyn:eq:model}\\
 \bar H(d)={}&-\sum_{i\ne j}\frac{\cos(E\chi_{ij})}{4\pi\sinh\chi_{ij}}
 \sigma_i^+\sigma_j^-.
 \label{eq:Hbar}
\end{align}
We use $\mathcal D[A]\rho=A\rho A^\dagger-\{A^\dagger A,\rho\}/2$ below.
The faithful stationary state and local excitation probability are
\begin{equation}
 \rho_\beta=\rho_{\rm ss}
 =\bigotimes_{i=1}^N\frac{\operatorname{diag}(1,q)}{1+q}
 =\frac{e^{-2\pi H_\eta}}{\Tr e^{-2\pi H_\eta}},\qquad
 u=\frac{q}{1+q}.
 \label{fn:dyn:eq:thermal}
\end{equation}
Exchange conserves excitation number, hence $[\bar H,\rho_\beta]=0$.
For every linear lowering channel $A$, $\rho_\beta A=q^{-1}A\rho_\beta$,
so $\mathcal D[A]\rho_\beta+q\mathcal D[A^\dagger]\rho_\beta=0$.
This proves stationarity. Uniqueness for sufficiently small positive $d$
is part of Theorem~\ref{fn:dyn:thm:full}.

\section{Numerical construction and control calculations}
\label{sec:numerics}

\subsection{Spectral projector and optimized preparation}

Density matrices are vectorized by column stacking; for $N=4$ the
Liouvillian is a $256\times256$ matrix. Denote its reduced eigenvalues by
$\lambda_k$, with $\bar\gamma_k=-\operatorname{Re}\lambda_k$; rates in
Rindler time are $\gamma_k=g^2\bar\gamma_k$.
For a diagonalizable generator, left and right eigenoperators obey
\begin{equation}
 \mathcal L_dR_k=\lambda_kR_k,\qquad
 \mathcal L_d^\dagger L_k=\lambda_k^*L_k,\qquad
 \Tr(L_k^\dagger R_{k'})=\delta_{kk'},\qquad
 c_k=\Tr[L_k^\dagger(\rho-\rho_\beta)].
\end{equation}
For the numerical modal amplitudes reported in the Letter,
we additionally impose
$\|R_k\|_{\rm HS}=[\Tr(R_k^\dagger R_k)]^{1/2}=1$.
For a fixed eigenoperator basis, this normalization,
together with biorthogonality, fixes the magnitude of
each modal coefficient; the remaining phase freedom
does not affect $|c_k|$.
The products $R_kc_k$ and the spectral projectors are
invariant under
$R_k\mapsto a_kR_k$ and
$L_k\mapsto L_k/a_k^*$, with $a_k\ne0$.
Then $\rho(s)-\rho_\beta=\sum_{k\ne0}c_ke^{\lambda_ks}R_k$.

The basis-independent cancellation condition is
$\mathcal P_{\rm slow}(\rho-\rho_\beta)=0$.
The projector is computed using an ordered complex Schur
decomposition, which does not require diagonalizability,
rather than relying solely on individual eigenvectors near
degeneracies. If $\mathcal L_d=UTU^\dagger$,
with the selected slow band ordered first, solve
\begin{equation}
 T=\begin{pmatrix}A&B\\0&C\end{pmatrix},\qquad
 AX-XC=-B,\qquad
 \mathcal P_{\rm slow}
 =U\begin{pmatrix}I&-X\\0&0\end{pmatrix}U^\dagger.
 \label{eq:schurprojector}
\end{equation}
The stationary mode is excluded from the selected band.
The spectra of $A$ and $C$ are disjoint, ensuring a unique
solution of the Sylvester equation.
The correction involving $X$ accounts for nonnormality;
the orthogonal Schur projector alone is generally not
the spectral projector.

For all $N\geq2$ define
\begin{equation}
 \rho_0=|10^{N-1}\rangle\langle10^{N-1}|,\qquad
 \rho_1=|1^N\rangle\langle1^N|,\qquad
 \rho_A(p)=(1-p)\rho_0+p\rho_1.
 \label{eq:family}
\end{equation}
With $y_\alpha=\mathcal P_{\rm slow}(\rho_\alpha-\rho_\beta)$
and $y_1\ne y_0$, the unconstrained Hilbert--Schmidt minimizer is
\begin{equation}
 p_\star=-\frac{\operatorname{Re}\Tr[(y_1-y_0)^\dagger y_0]}
                    {\|y_1-y_0\|_{\HS}^2},\qquad
 \mathcal P_{\rm slow}[\rho_A(p)-\rho_\beta]
 =y_0+p(y_1-y_0).
 \label{eq:numericalpstar}
\end{equation}
We check $0<p_\star<1$ and a vanishing projected residual.
At the Letter's reference point
$(N,E,d)=(4,0.424264,0.01)$, the slow band is
one-dimensional, $p_\star=0.7715429438$ (rounded), and the reported
slow-band residual is of order $10^{-14}$ using the full-precision root.

The trace-distance comparison uses
\begin{equation}
 \rho_B^{\rm num}=(1-r)\rho_\beta+r|F_2\rangle\langle F_2|,
 \quad
 |F_2\rangle=\tfrac12
 (|1000\rangle-|0100\rangle+|0010\rangle-|0001\rangle),
 \quad r=0.6499877261.
 \label{eq:Bnum}
\end{equation}
The reduced rates are
$\bar\gamma_{\rm slow}=5.06156\times10^{-5}$ and
$\bar\gamma_{k_*}=8.22472\times10^{-4}$, yielding
$\mathcal R_4=\bar\gamma_{k_*}/\bar\gamma_{\rm slow}=16.2494$;
$k_*$ is the slowest remaining mode carrying nonzero weight in the ideal
optimized state. The spectral cancellation is exact for the theoretical
root; finite numerical or preparation errors can leave a late slow tail.

The same procedure is applied across the Letter's grid,
$0.28\leq E\leq0.58$ and $0.006\leq d\leq0.018$, to obtain
$p_\star(E,d)$ and $\mathcal R(E,d)$ with physical preparations
$0<p_\star<1$. The theorem establishes existence for sufficiently small
$d$ at each fixed $N,E$; the scan separately describes the displayed
finite parameter region.

\subsection{Local-rate-preserving control}
At the reference point we remove cross rates and exchange,
$\bar\Gamma^{\downarrow,\uparrow}_{ij,\rm loc}
=\delta_{ij}\bar\Gamma^{\downarrow,\uparrow}_{ii,U}$ and
$\bar J_{ij,\rm loc}=0$ for $i\ne j$, retaining
\begin{equation}
 \kappa_\downarrow=0.0725709194,\quad
 \kappa_\uparrow=0.00504720665,\quad
 \kappa_\Sigma=0.0776181261,\quad u=0.06502613.
\end{equation}
The observables $O_j=\sigma_j^+\sigma_j^--uI$ obey
$\mathcal L_{\rm loc}^\dagger O_j=-\kappa_\Sigma O_j$. On the family
in Eq.~\eqref{eq:family},
\begin{equation}
 (\langle O_1\rangle_A,\ldots,\langle O_4\rangle_A)
 =(1-u,p-u,p-u,p-u).
 \label{eq:localobstruction}
\end{equation}
The first entry is nonzero for every $p$, so the full accessible slow band
cannot be cancelled within this family. This band is degenerate, and
testing a single eigenvector would be insufficient.
For precisely the same initial pair, without retuning $p$ or $r$,
\begin{equation}
 (D_A,D_B)_{\rm coll}(10)=(0.515016,0.599847),\qquad
 (D_A,D_B)_{\rm loc}(10)=(0.611254,0.277484),
 \quad D_X(s)=\tfrac12\|\rho_X(s)-\rho_\beta\|_1.
\end{equation}
The collective crossing is at $s_\times\simeq8.3674$; the local pair does
not cross in the checked interval $0\leq s\leq400$. A supplementary grid
of 201 preparation values and 601 times in $0\leq s\leq150$ found no
initially farther family member crossing $\rho_B^{\rm num}$. This finite
grid is not a proof against every possible transient crossing; the exact
cancellation obstruction follows from Eq.~\eqref{eq:localobstruction}.
Since $\kappa_\Sigma>\bar\gamma_{k_*}$, the collective effect is a reversal
of the ordering of two states, not an absolute speedup over local decay.

\subsection{Common thermal-field benchmark}
\label{sec:thermal_corollary_benchmark}

\begin{figure}[htbp]
\centering
\includegraphics[width=0.90\textwidth]
{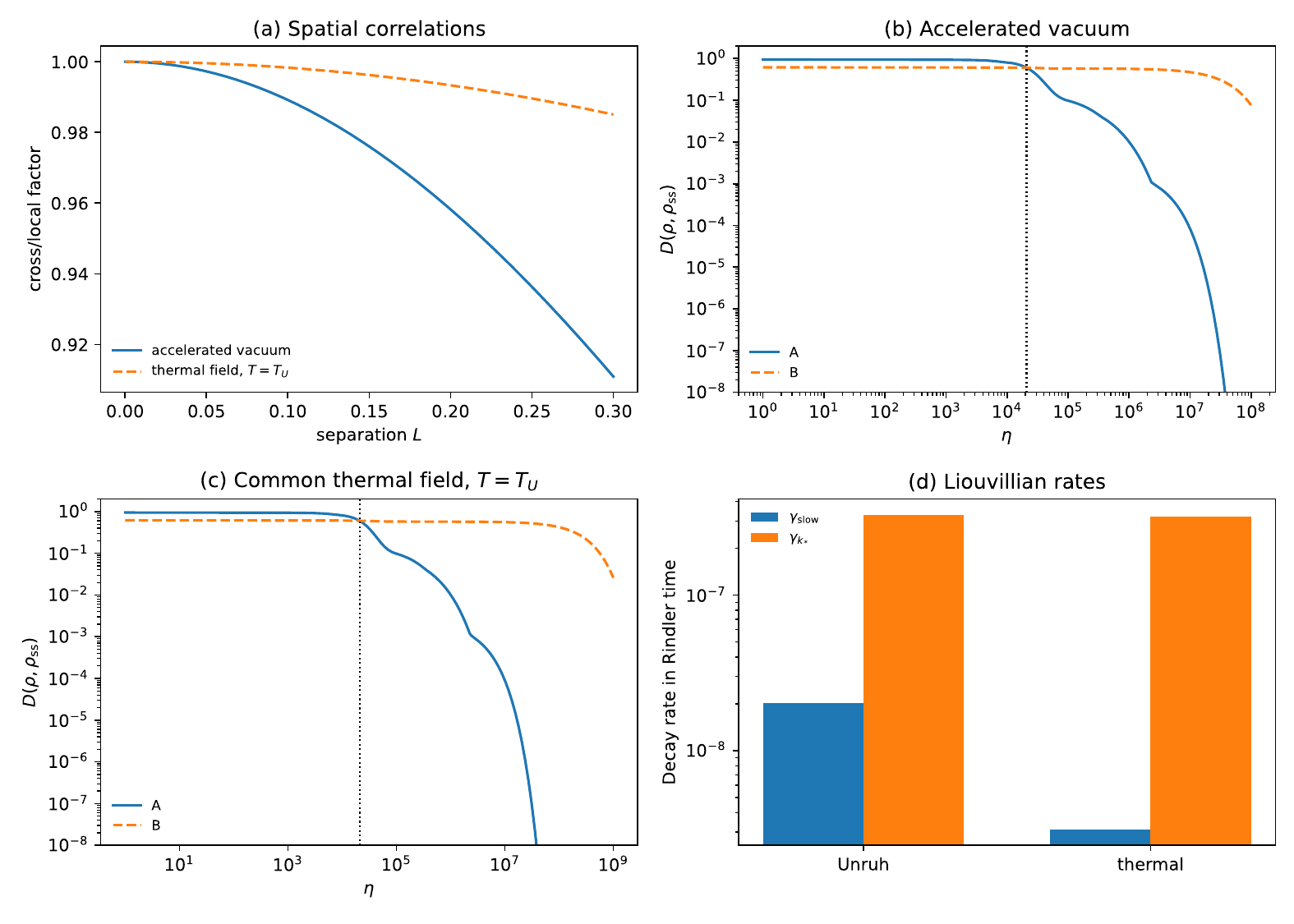}
\caption{Accelerated vacuum versus a common thermal field at $T=T_U$.
(a) Spatial cross-correlation factors ($\omega_0=1$); the local limits
coincide. (b),(c) Optimized trace-distance relaxation in the two environments.
(d) Slow and first accessible faster rates at the reference point.
The time axes $\eta$ and unreduced rates use $g=0.02$ only to set the scale;
Table~\ref{tab:thermal} is independent of that choice.}
\label{fig:thermal}
\end{figure}

For inertial detectors in a common field at $T=T_U$, the thermal
correlation function is
\begin{equation}
 G^+_\beta(\Delta t,R_{ij})=-\frac1{4\pi^2}
 \sum_{n=-\infty}^{\infty}
 \frac1{(\Delta t-in\beta-i\epsilon)^2-R_{ij}^2},
 \quad R_{ij}=|i-j|L,\quad\beta=\frac{2\pi}{a}.
 \label{eq:SM_thermal_correlation}
\end{equation}
Using $\eta=at$ gives $\Gamma^{(\eta)}=\Gamma^{(t)}/a$. The prefactors
$a_\downarrow,a_\uparrow$ and stationary state therefore coincide with
the accelerated case, but the spatial factors and exchange are
\begin{equation}
 F^{\rm th}_{ij}=\frac{\sin(\omega_0R_{ij})}{\omega_0R_{ij}},
 \qquad
 \bar J^{\rm th}_{ij}
 =-\frac{E\cos(\omega_0R_{ij})}{4\pi\omega_0R_{ij}}
 \quad(i\ne j),
 \qquad F^{\rm th}_{ii}=1.
 \label{eq:SM_thermal_kernel_exchange}
\end{equation}
We optimize $p_\star$ independently for each generator while keeping the
family and reference state fixed. Thus this comparison concerns their
optimized spectral structures, not identical optimized initial states.
Table~\ref{tab:thermal} and Fig.~\ref{fig:thermal} show that both baths
support the effect. Equal local temperatures do not imply equal collective
generators or slow relaxation scales.

The small-separation comparison can also be obtained analytically.
For fixed finite $N$ and fixed $E>0$,
Eqs.~\eqref{eq:rates} and~\eqref{eq:SM_thermal_kernel_exchange} give
\begin{equation}
 F_{ij}^{U}
 =
 1-\frac{1+E^{-2}}{6}(i-j)^2d^2+O(d^4),
 \qquad
 F_{ij}^{\mathrm{th}}
 =
 1-\frac{1}{6}(i-j)^2d^2+O(d^4).
 \label{eq:SM_kernel_comparison}
\end{equation}
Both environments have the same local rates, stationary state,
and equilibrium reflection-sector weights. Their coherent exchange
Hamiltonians also share the leading term $H_{-1}/d$, with $H_{-1}$
defined in Eq.~\eqref{fn:dyn:eq:commutant}; their difference is
$O(d)$ and preserves reflection. The fast-complement estimates and
the sector reduction therefore apply to both generators, as detailed
in Sec.~\ref{sec:thermal_corollary}. In particular, the leading
intersector transfer rate is proportional to the quadratic coefficient
of the spatial kernel. Consequently,
\begin{equation}
 \bar\gamma_{\mathrm{slow}}^{\mathrm{th}}(d)
 =
 \kappa_{N,E}^{\mathrm{th}}d^2+O(d^4),
 \qquad
 \kappa_{N,E}^{\mathrm{th}}
 =
 \frac{N(N^2-1)E}
 {36\pi\sinh(2\pi E)(1-t_N^2)},
 \label{eq:SM_thermal_slow_rate}
\end{equation}
where $t_N=\operatorname{Tr}(\Pi_N\rho_\beta)$.
Comparing with the accelerated-vacuum coefficient in
Eq.~\eqref{fn:dyn:eq:kappa} yields
\begin{equation}
 \lim_{d\to0^+}
 \frac{\bar\gamma_{\mathrm{slow}}^{U}(d)}
      {\bar\gamma_{\mathrm{slow}}^{\mathrm{th}}(d)}
 =
 1+E^{-2}.
 \label{eq:SM_slow_rate_ratio}
\end{equation}
This limit holds for each fixed finite $N$ and fixed $E>0$.
It compares slow relaxation rates, not strong-Mpemba enhancement
factors, and does not assert an exact ratio at finite separation
or a validity interval uniform in $N$.
The fixed-$N$ analytical extension and small-$d$ rate comparison
are proved in Sec.~\ref{sec:thermal_corollary}.

\begin{table}[htbp]
\centering
\caption{Reference-point comparison at $N=4$, $E=0.424264$, $d=0.01$.
All rates are reduced by $g^2$.}
\label{tab:thermal}
\begin{tabular}{lrrrr}
\toprule
Environment & $p_\star$ & $\bar\gamma_{\rm slow}$
& $\bar\gamma_{k_*}$ & $\mathcal R$\\
\midrule
Accelerated vacuum & 0.771543 & $5.06156\times10^{-5}$
& $8.22472\times10^{-4}$ & 16.25\\
Common thermal field & 0.771588 & $7.76739\times10^{-6}$
& $8.02963\times10^{-4}$ & 103.38\\
\bottomrule
\end{tabular}
\end{table}

\FloatBarrier
\subsection{Spatial-profile checks within the secular generator}
\label{SMfp:section}

\paragraph{Profile and interaction.}
Write $x=aL=d/E$ and $y=a\ell<1$. Introduce dimensionless
radial displacement $u=a\xi-1$ and transverse displacement
$\mathbf v=a(\mathbf r_\perp-\mathbf r_{\perp,i})$.
The normalized, boost-stationary profile is
\begin{align}
 f_y(u,\mathbf v)&=b_y(u)\,g_y(\mathbf v),\qquad
 g_y(\mathbf v)=\frac{e^{-|\mathbf v|^2/(2y^2)}}{2\pi y^2},
 \label{SMfp:profile}\\
 b_y(u)&=\frac{1}{yZ_b}
 \begin{cases}
 e^{-1/[1-(u/y)^2]},& |u|<y,\\
 0,& |u|\ge y,
 \end{cases}
 \qquad
 Z_b=\int_{-1}^{1}e^{-1/(1-t^2)}\,dt.
 \label{SMfp:bump}
\end{align}
Thus $\int du\,d^2v\,f_y=1$. The physical spatial density is
$a^3f_y$. The radial support is $a^{-1}-\ell<\xi<a^{-1}+\ell$;
$\ell$ is its half-width, not its root-mean-square width.
Each transverse Cartesian coordinate has standard deviation $\ell$.
The transverse Gaussian has infinite tails, so $\ell/L$ is not a
strict no-overlap criterion. The radial support remains inside the
right Rindler wedge.

With $\mathbf z_i=((i-1)x,0)$, the interaction is
\begin{equation}
 H_I(\eta)=g\sum_i\mu_i(\eta)\Phi_i^{(y)}(\eta),\qquad
 \Phi_i^{(y)}(\eta)=\int du\,d^2v\, f_y(u,\mathbf v)
 \frac{1+u}{a}\,
 \phi\!\left(\eta,\xi=\frac{1+u}{a},
                  \mathbf r_\perp=\frac{\mathbf z_i+\mathbf v}{a}\right).
 \label{SMfp:interaction}
\end{equation}
The factor $(1+u)/a=\xi$ is part of the coupling convention, and
$\mu_i(\eta)$ has the common Rindler frequency $E$ throughout the profile.
Its pointlike limit is the interaction used in the Letter.

\paragraph{One correlation for dissipation and exchange.}
For two independently sampled profile points, let
\begin{equation}
 \chi=2\operatorname{arsinh}
 \frac{\sqrt{(u-u')^2+
 |\mathbf z_i-\mathbf z_j+\mathbf v-\mathbf v'|^2}}
 {2\sqrt{(1+u)(1+u')}}.
 \label{SMfp:chi}
\end{equation}
Let $\langle\cdot\rangle_y$ denote averaging with the product measure
$f_y(u,\mathbf v)f_y(u',\mathbf v')\,du\,du'\,d^2v\,d^2v'$.
Including the two coupling factors in Eq.~\eqref{SMfp:interaction},
the common correlation is
\begin{equation}
 W^{(y)}_{ij}(v)=
 -\frac{1}{8\pi^2}\left\langle
 \frac{1}{\cosh(v-i0)-\cosh\chi}
 \right\rangle_y .
 \label{SMfp:correlation}
\end{equation}
Here the argument $v$ is a Rindler-time difference, distinct from
the transverse vector $\mathbf v$. In reduced time $s=g^2\eta$,
\begin{equation}
 \bar\Gamma^\downarrow_{ij}
 =a_\downarrow\left\langle
 \frac{\sin(E\chi)}{E\sinh\chi}\right\rangle_y,
 \qquad
 \bar\Gamma^\uparrow_{ij}=q\bar\Gamma^\downarrow_{ij},
 \qquad q=e^{-2\pi E}.
 \label{SMfp:rates}
\end{equation}
Unlike the pointlike kernel, the averaged diagonal factor need not
be unity. The change in the local response is retained.
The sum of the positive- and negative-frequency principal-value
transforms of the same correlation gives, for distinct detectors,
\begin{equation}
 \bar J^{(y)}_{ij}
 =-\frac{1}{4\pi}
 \left\langle\frac{\cos(E\chi)}{\sinh\chi}\right\rangle_y,
 \qquad i\ne j.
 \label{SMfp:exchange}
\end{equation}
Common local frequency shifts are absorbed into the renormalized gap.
The Hamiltonian and dissipator are then assembled as in the pointlike
calculation. The matrix detailed-balance relation and excitation-number
conservation retain the same product thermal stationary state $\rho_\beta$.

\paragraph{Numerical integration.}
The dissipative coefficients are evaluated by the equivalent Rindler-mode
integral, where $r=|i-j|x$ and $k$ is dimensionless transverse momentum:
\begin{equation}
 I_y(E,r)=\int_0^\infty k\,dk\,J_0(kr)e^{-y^2k^2}
 \left[\int_{-y}^{y}du\,b_y(u)(1+u)
 K_{iE}\bigl(k(1+u)\bigr)\right]^2,
 \qquad
 \bar\Gamma^\downarrow_{ij}
 =a_\downarrow\frac{I_y(E,r)}{I_0(E,0)}.
 \label{SMfp:modeintegral}
\end{equation}
Here $I_0(E,0)=\pi E/[2\sinh(\pi E)]$.
The code separately records the local factor $I_y(E,0)/I_0(E,0)$
and the normalized cross factor $I_y(E,r)/I_y(E,0)$.
It evaluates $K_{iE}(z)=\int_0^\infty e^{-z\cosh t}\cos(Et)\,dt$
using Gauss--Legendre quadrature on $0\le t\le28$, and uses
Gauss--Legendre quadrature for the radial integral. Composite Simpson
integration is applied on a uniform grid in $\log k$, with
$10^{-9}\le k\le80$. The nominal calculation uses respectively
$2400$, $190$, and $28$ nodes for the $\log k$, Bessel-representation,
and radial integrals. The separate high-acceleration calculations
below use $1400$, $150$, and $22$ nodes.

Equation~\eqref{SMfp:exchange} is evaluated by scrambled Sobol sampling.
After integrating out the common transverse coordinate, only four
independent random variables remain: two radial displacements and
two transverse differences with standard deviation $\sqrt{2}y$.
The radial inverse cumulative distribution is tabulated on $5001$
uniform nodes in $u/y\in[-1,1]$ and linearly interpolated.
The nominal calculation uses eight scrambles, with seeds $813$--$820$
and $2^{17}$ points per scramble; the separate high-acceleration
calculation uses six scrambles with $2^{14}$ points each.
The reported sampling error is the standard deviation of the
replicate means divided by the square root of their number.

\paragraph{Preparations and separate high-acceleration example.}
For $N=4,5,6$, we use the preparation family already defined above,
$\rho_A(p)=(1-p)|10^{N-1}\rangle\langle10^{N-1}|
+p|1^N\rangle\langle1^N|$, and the reference state
\begin{equation}
 \rho_B^{(N)}=(1-r_B)\rho_\beta+r_B|F_N\rangle\langle F_N|,
 \qquad
 |F_N\rangle=\frac{1}{\sqrt N}\sum_{j=1}^N(-1)^{j-1}|1_j\rangle,
 \qquad r_B=0.6499877261.
 \label{SMfp:reference}
\end{equation}
Here $|1_j\rangle$ has only detector $j$ excited. The results at
$E=7.957747\times10^{-6}$ and $aL=0.05$ are listed in
Table~\ref{SMfp:highacc}. In every case $0<p_\star<1$,
$D_A(0)>D_B(0)$, and the optimized slow-mode amplitude is at numerical
roundoff level. The initial states lie in the invariant block of zero
excitation-number difference. Its slowest mode is nondegenerate and
separated from the next rate in all six calculations. This removes the
entire slow band accessible to these preparations; it does not identify
the slowest mode across all coherence blocks of the full Liouvillian.

\begin{table}[htbp]
\centering
\caption{Separate high-acceleration finite-profile example. Crossings
are in reduced time $s=g^2\eta$; $\mathcal R$ is the ratio of the next
accessible rate to the isolated slow rate.}
\label{SMfp:highacc}
\begin{tabular}{cccc}
\hline
$N$ & $a\ell$ & $s_\times$ & $\mathcal R$\\
\hline
4&0.0025&14.2058&2.864\\
5&0.0025&23.0599&6.584\\
6&0.0025&25.2037&1.922\\
4&0.0050&14.1737&2.914\\
5&0.0050&22.9847&6.666\\
6&0.0050&25.1266&1.954\\
\hline
\end{tabular}
\end{table}

\paragraph{Reference-point results and refinement.}
At $N=4$, $E=0.424264$, and $d=0.01$, the center separation is
$x=d/E=0.02357022986$. Table~\ref{SMfp:nominal} documents the
pointlike limit and the two nonzero widths considered here. The reference
state is fixed, while $p_\star$ is optimized separately for each generator.
The initial distances are $D_A(0)\simeq0.946834$ and
$D_B(0)\simeq0.615442$ in all three cases. The next rate listed here
belongs to a mode with nonzero weight in the optimized state.

\begin{table}[htbp]
\centering\small
\setlength{\tabcolsep}{4pt}
\caption{Finite-profile checks at the Letter's reference geometry.
All rates are reduced by $g^2$. Values from the original nominal
calculation are rounded conservatively. The last column is the
Hilbert--Schmidt norm of the slow spectral projection, evaluated
with the unrounded optimized preparation in the discretized generator.}
\label{SMfp:nominal}
\begin{tabular}{ccccccc}
\hline
$a\ell$ & $p_\star$ & $s_\times$ & $\bar\gamma_{\rm slow}$ &
$\bar\gamma_{\rm next}$ & $\mathcal R$ & $\epsilon_{\rm proj}$\\
\hline
0 & 0.7715429 & 8.36739 & $5.06156\times10^{-5}$ &
$8.22\times10^{-4}$ & 16.25 & $<5\times10^{-14}$\\
0.001 & 0.7715445 & 8.36755 & $5.06158\times10^{-5}$ &
$8.26\times10^{-4}$ & 16.32 & $<5\times10^{-14}$\\
0.0025 & 0.7715532 & 8.36767 & $5.06166\times10^{-5}$ &
$8.46\times10^{-4}$ & 16.71 & $<5\times10^{-14}$\\
\hline
\end{tabular}
\end{table}

For a simple slow eigenvalue, the residual is computed as
\begin{equation}
 \epsilon_{\rm proj}=
 \left\|R_s\frac{\operatorname{Tr}[L_s^\dagger
 (\rho_A-\rho_\beta)]}{\operatorname{Tr}(L_s^\dagger R_s)}
 \right\|_{\rm HS}.
 \label{SMfp:residual}
\end{equation}
This expression is independent of the separate eigenoperator scales.
Right eigenoperators are normalized to unit Hilbert--Schmidt norm
when reporting scalar modal amplitudes.

For both nonzero widths at the reference point, increasing the
quadrature to $(3200,260,56)$ nodes changed the local factor by less
than $2.1\times10^{-13}$ and each normalized cross factor at separations
$x$, $2x$, and $3x$ by less than $10^{-15}$.
Extending the integration ranges to $10^{-11}\le k\le100$ and
$0\le t\le32$, using $(3600,300,56)$ nodes, changed these factors
by less than $10^{-15}$ relative to the preceding refinement.
These are observed numerical differences, not rigorous error bounds.

Doubling the Sobol sample to $2^{18}$ points per scramble, together
with the first quadrature refinement, changed $p_\star$ by less than
$3\times10^{-9}$, $s_\times$ by less than $7\times10^{-8}$,
$\bar\gamma_{\rm slow}$ by less than $1.2\times10^{-8}$ relatively,
and $\bar\gamma_{\rm next}$ by less than $8\times10^{-6}$ relatively.
At the original resolution, the nearest-neighbor exchange estimates
for $y=0.001$ and $0.0025$ are respectively $-3.38093293$ and
$-3.40923530$, with replicate standard errors $2.2\times10^{-7}$
and $3.7\times10^{-6}$. At doubled resolution they are
$-3.38093285$ and $-3.40922827$, with standard errors
$9.0\times10^{-8}$ and $2.0\times10^{-6}$. The preparation is
reoptimized for each discretization. Thus an algebraic residual
near roundoff does not bound the quadrature error in the continuum
slow projector or the sensitivity to preparation errors.

These calculations support persistence of the crossing and accessible
slow-mode cancellation within the chosen spatially smeared,
second-order secular generator. The pointlike rates and ratio quoted
for the Letter's main figures are not assigned to the smeared generators.
Microscopic finite-coupling errors are not bounded by these integration
checks; their interpretation remains as stated in the following subsection.

\subsection{Weak-coupling interpretation and order of limits}
The secular weak-coupling construction uses the van Hove scaling
$g\to0$, $\eta\to\infty$ with $s=g^2\eta$ fixed, at fixed $E,d$.
The rates retain the full dependence on $E$; no high-temperature expansion
is used. The Letter's spectra and parameter scan belong to the displayed
pointlike generator. Spatial smearing tests persistence within the
second-order secular framework, rather than bounding its microscopic
error at finite coupling.
For a representative finite coupling, separation-of-scale diagnostics are
\begin{equation}
 \frac{\lambda_{\max}(\Gamma^{\uparrow,\downarrow})}{E}\ll1,\qquad
 \frac{\max_{i\ne j}|J_{ij}|}{E}\ll1,\qquad
 \epsilon_{\rm dark}=\frac{g^2}{(aL)^2}=\frac{g^2E^2}{d^2}\ll1.
 \label{eq:diagnostics}
\end{equation}
The last ratio is a heuristic geometric diagnostic, not an error bound.
At the reference geometry and $g=0.002$, these give
$\lambda_{\max}(\Gamma^\downarrow)/E\simeq2.74\times10^{-6}$,
$\max|J_{ij}|/E\simeq3.18\times10^{-5}$ and
$\epsilon_{\rm dark}=7.2\times10^{-3}$.
Profile-specific memory and higher-order corrections must still be
controlled relative to the small slow rate to establish a microscopic
finite-coupling regime. Here $aL=d/E\simeq0.02357$ and
$\pi E/(aL)^2\simeq2.40\times10^3$, well outside the small-value window
of the separate high-acceleration profile example. The reduced rates,
$p_\star$ and $\mathcal R$ of the second-order generator are independent
of the representative $g$.

The finite-$N$ theorem below concerns the pointlike secular generator:
its small-$d$ limit is taken at the generator level after the weak-coupling
construction at fixed geometry. It does not assert a uniform microscopic
approximation in a simultaneous $g,d\to0$ limit.

\subsection{Large-$N$ residual of the leading parity candidate}
\label{sec:largeNparity}

We assess the leading parity candidate
\begin{equation}
 Q_N=\Pi-t_N I,\qquad
 t_N=\operatorname{Tr}(\rho_\beta\Pi)
 =\left[\frac{1+q^2}{(1+q)^2}\right]^{\lfloor N/2\rfloor},
 \qquad q=e^{-2\pi E},
\end{equation}
using the full finite-separation spatial kernel of the pointlike
generator. We use the thermal inner product and norm
\begin{equation}
 \langle A,B\rangle_\beta
 =\operatorname{Tr}(\rho_\beta A^\dagger B),
 \qquad
 \|A\|_\beta=\sqrt{\langle A,A\rangle_\beta},
\end{equation}
for which $\|Q_N\|_\beta^2=1-t_N^2$. Define
\begin{equation}
 \gamma_\Pi
 =-\frac{\langle Q_N,\mathcal L_d^\dagger Q_N\rangle_\beta}
          {1-t_N^2},
 \qquad
 r_\Pi
 =\frac{\|\mathcal L_d^\dagger Q_N+\gamma_\Pi Q_N\|_\beta}
          {\sqrt{1-t_N^2}}.
 \label{eq:parity_rayleigh_residual}
\end{equation}
Here $\gamma_\Pi$ is a Rayleigh quotient, not an independently
computed slow eigenvalue, and $r_\Pi$ measures the failure of
$Q_N$ to satisfy the corresponding eigenoperator equation.

This calculation requires only $N\times N$ spatial matrices.
Let $\mathsf R_{ij}=\delta_{i,N+1-j}$ denote spatial reflection
and $P_{\rm o}=(I-\mathsf R)/2$. Only the odd spatial block
$F_{\rm o}=P_{\rm o}FP_{\rm o}$ contributes to
$\mathcal L_d^\dagger Q_N$; the reflection-invariant Hamiltonian
commutes with $Q_N$. The thermal averages in
Eq.~\eqref{eq:parity_rayleigh_residual} can therefore be evaluated
without constructing many-body operators or diagonalizing the
Liouvillian, with $O(N^2)$ work and storage per parameter point.

\begin{figure}[htbp]
 \centering
 \includegraphics[width=\textwidth]
 {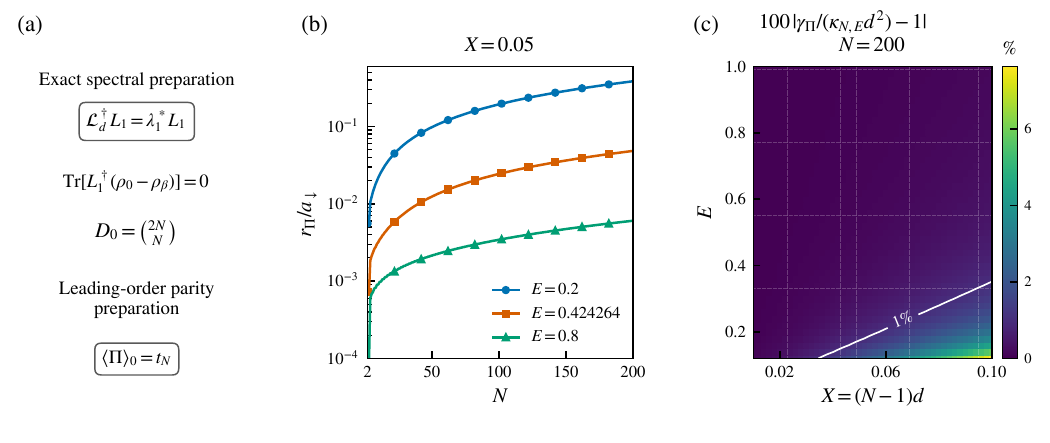}
 \caption{
 Large-$N$ evaluation of the leading parity candidate. 
 (a) The exact spectral preparation condition and its leading
 parity approximation, $\langle\Pi\rangle_0=t_N$.
 The excitation-number-conserving operator block has dimension
 $D_0=\binom{2N}{N}$, whereas the parity-residual calculation
 uses only spatial matrices.
 (b) Thermal-norm residual
 $r_\Pi/a_\downarrow$, defined in
 Eq.~\eqref{eq:parity_rayleigh_residual}, for every integer
 $2\leq N\leq200$ at $X=(N-1)d=0.05$ and the indicated values
 of $E$. Solid lines connect results at consecutive integer values of $N$;
symbols mark selected points from the same calculation.
 (c) Percentage discrepancy
 $100|\gamma_\Pi/(\kappa_{N,E}d^2)-1|$ at $N=200$;
 the contour marks $1\%$.
 The comparison uses the Rayleigh quotient of the parity
 candidate, not an independently computed Liouvillian
 eigenvalue. No Liouvillian eigensolver is used in these scans.
 }
 \label{fig:largeNparity}
\end{figure}

Figure~\ref{fig:largeNparity} reports the scan over every integer
$2\leq N\leq200$ at fixed dimensionless chain extent
$X=(N-1)d=0.05$. A separate map at $N=200$ compares
$\gamma_\Pi$ with $\kappa_{N,E}d^2$, where
$\kappa_{N,E}$ is given in Eq.~\eqref{fn:dyn:eq:kappa}.
On the sampled map points satisfying $E\geq0.2$ and $X\leq0.05$,
the relative discrepancy is below $0.76\%$.

This agreement concerns the finite-separation Rayleigh quotient
and its leading small-$d$ expression. It does not establish the
same accuracy for the true slow eigenvalue. The residual remains
nonzero and grows with $N$ along the fixed-$X$ scans. For example,
at $N=200$, $E=0.424264$, and $X=0.05$, we find
$r_\Pi/a_\downarrow\simeq0.04837$ and
$r_\Pi/\gamma_\Pi\simeq2.026$, with
$a_\downarrow=E/[2\pi(1-e^{-2\pi E})]$.
Thus the calculation tests the leading parity candidate rather
than certifying exact slow-mode cancellation at finite separation.
It neither establishes a uniform-in-$N$ validity interval nor
replaces the finite-$N$ spectral construction.

\section{Reflection commutant for arbitrary finite detector number}
\label{sec:fn_alg}
\subsection{Statement and elementary algebra}
On $(\mathbb C^2)^{\otimes N}$ define
\begin{equation}
 B_N=\sum_{i<j}\frac{\sigma_i^+\sigma_j^-+\sigma_i^-\sigma_j^+}{j-i},
 \qquad J_\pm=\sum_i\sigma_i^\pm,
 \qquad \Pi_N|s_1\cdots s_N\rangle=|s_N\cdots s_1\rangle.
\end{equation}
\begin{fntheorem}\label{fn:alg:thm:main}
For every integer $N\geq2$,
\begin{equation}
 \{B_N,J_+,J_-\}'=\operatorname{span}_{\mathbb C}\{\fnIdentity,\Pi_N\}.
 \label{fn:alg:eq:goal}
\end{equation}
\end{fntheorem}
Let $\fnAlg$ be the unital complex associative algebra generated by these operators. It is a finite-dimensional $*$-algebra. Collective rotations belong to $\fnAlg$, as do their conjugations of its elements. Since
$B_N=\frac12\sum_{i<j}(X_iX_j+Y_iY_j)/(j-i)$, rotating the coordinate axes gives
\begin{equation}
 D=\sum_{i<j}\frac{Z_iZ_j}{j-i}\in\fnAlg.
 \label{fn:alg:eq:D}
\end{equation}
Indeed, the sum of the $yz$ and $zx$ versions of $B_N$, minus its $xy$ version, equals $D$. To match the detector convention explicitly, take
$\sigma^-=|0\rangle\langle1|$, $\sigma^+=(\sigma^-)^\dagger$,
$X=\sigma^++\sigma^-$, $Y=-i(\sigma^+-\sigma^-)$, and
$Z=[\sigma^+,\sigma^-]$. Thus $\sigma^\pm=(X\pm iY)/2$ and
$[Z,\sigma^\pm]=\pm2\sigma^\pm$. Overall nonzero normalizations of generators do not affect their commutant.

\subsection{Arithmetic separation of a shell}
For a consecutive interval $I$ of $L$ sites write
\begin{equation}
 S_I^\alpha=\sum_{i\in I}\sigma_i^\alpha,\qquad
 D_I=\sum_{i<j\in I}\frac{Z_iZ_j}{j-i},
 \quad \alpha=x,y,z,+,-.
\end{equation}
\begin{fnlemma}[Central reduction]\label{fn:alg:lem:peel}
Suppose $S_I^{x,y,z}\in\fnAlg$ and $K_I=D_I+R\in\fnAlg$, where $R$ acts only outside $I$. If $L\geq3$, a smaller, nonempty central interval $I'\subset I$ of the same parity of length satisfies the same assumptions, with a new outside operator $R'$. Consequently $\fnAlg$ contains the collective spin of the central single site for odd $N$, and of the central pair for even $N$.
\end{fnlemma}
\begin{fnproof}
Choose a prime $p$ with $L/2<p<L$. For $L=3$ take $p=2$; for $L\geq4$ its existence follows from Bertrand's theorem, applied to $\lfloor L/2\rfloor$. Label the sites of $I$ by $1,\ldots,L$ and set
\begin{equation}
 A=\{1,\ldots,L-p\}\cup\{p+1,\ldots,L\},
 \qquad I'=\{L-p+1,\ldots,p\}.
\end{equation}
The shell $A$ consists exactly of sites having a partner at distance $p$. Every such site has exactly one such partner; no site of $I'$ has one.

The matrix elements of $\sigma_i^+$ in the computational basis have adjoint frequencies under $D_I$ equal to
\begin{equation}
 2f_i(z),\qquad f_i(z)=\sum_{j\in I\setminus\{i\}}\frac{z_j}{|i-j|},\qquad z_j\in\{-1,1\}.
 \label{fn:alg:eq:freq}
\end{equation}
Let $M=\operatorname{lcm}(1,\ldots,L-1)$. Write $v_p(n)$ for the exponent of the prime $p$ in the integer $n$.
Since $2p>L$, the only multiple of $p$ among $1,\ldots,L-1$ is $p$
itself. In particular $v_p(M)=1$. Every distance $k\ne p$ in this range
satisfies $v_p(k)=0$, and $k$ divides $M$, so
\begin{equation}
 v_p(M/k)=1\quad(k\ne p),\qquad v_p(M/p)=0.
\end{equation}
Equivalently, $p\mid M/k$ for all $k\ne p$, whereas $p\nmid M/p$.
There is exactly one distance-$p$ term at each shell site, so it cannot
cancel against another distance-$p$ term. Thus
\begin{align}
 M f_i(z)&\equiv \pm M/p\not\equiv0\pmod p &&(i\in A),\\
 M f_i(z)&\equiv0\pmod p &&(i\in I').
\end{align}
In particular, the finite sets of frequencies for shell and interior sites are disjoint. Let $\Omega_A$ and $\Omega_{I'}$ be the sets of distinct shell and interior
frequencies $2f_i(z)$, and $\Omega=\Omega_A\cup\Omega_{I'}$.
Disjointness of the $f_i$ values implies disjointness of their doubles
as real numbers, including when $p=2$; no division by two modulo $p$
is used. Repeated frequencies within either set are listed only once.
An explicit interpolation is
\begin{equation}
 h(z)=\sum_{\omega\in\Omega_A}
 \prod_{\nu\in\Omega\setminus\{\omega\}}\frac{z-\nu}{\omega-\nu}.
\end{equation}
All denominators are nonzero because the nodes are distinct. This is a real polynomial equal to one on shell frequencies and zero on
interior frequencies. Modular arithmetic establishes disjointness only;
the polynomial acts on the actual real frequencies, not on residue classes. Because $R$ acts outside $I$, repeated commutators with $K_I$ on operators supported in $I$ agree with those with $D_I$. Therefore
\begin{equation}
 h(\fnAd_{K_I})S_I^+=S_A^+\in\fnAlg.
\end{equation}
Taking adjoints and commutators gives $S_A^{x,y,z}\in\fnAlg$, and subtracting from $S_I^{x,y,z}$ gives $S_{I'}^{x,y,z}\in\fnAlg$.

Average $K_I$ over collective $SU(2)$ rotations on $A$. The cross terms between $A$ and $I'$ average to zero, while $Z_iZ_j$ with $i,j\in A$ averages to $\boldsymbol\sigma_i\cdot\boldsymbol\sigma_j/3$. Hence the average is
\begin{equation}
 K_{I'}=D_{I'}+
 \underbrace{R+\frac13\sum_{i<j\in A}
 \frac{\boldsymbol\sigma_i\cdot\boldsymbol\sigma_j}{j-i}}_{R'\text{ supported outside }I'}\in\fnAlg.
\end{equation}
The average belongs to $\fnAlg$: all rotated operators belong to this finite-dimensional linear space, which is closed. The new length is $2p-L$, strictly between zero and $L$ and with the same parity. Starting with $I=\{1,\ldots,N\}$, $K_I=D$, repeated reduction ends at length one or two.
\end{fnproof}

\subsection{A lemma for independently resolved reflection pairs}
Let $r_i=N+1-i$ and group sites into reflection pairs $\{i,r_i\}$. Their swaps are denoted $F_i$, with triplet/singlet projectors $P_{i,\pm}=(\fnIdentity\pm F_i)/2$.
\begin{fnlemma}\label{fn:alg:lem:pairs}
On any collection of reflection pairs of this chain, an operator commuting with the collective spin of each pair and with the exchange Hamiltonian restricted to those sites is a linear combination of the identity and the product of their swaps. For one pair this follows already from its collective spin.
\end{fnlemma}
\begin{fnproof}
The collective representation on each pair is the multiplicity-free sum of spin one and spin zero. Commuting with the independent pair rotations therefore forces the operator to have the form
\begin{equation}
 X=\sum_{\varepsilon\in\{+,-\}^r}x_\varepsilon
 \bigotimes_i P_{i,\varepsilon_i}.
\end{equation}
For two distinct pairs $i,j$, define $S_i^\pm=\sigma_i^\pm+\sigma_{r_i}^\pm$ and $A_i^\pm=\sigma_i^\pm-\sigma_{r_i}^\pm$. Their exchange interaction is
\begin{equation}
 \frac{a+b}{2}(S_i^+S_j^-+S_i^-S_j^+)
 +\frac{a-b}{2}(A_i^+A_j^-+A_i^-A_j^+),
 \quad a=\frac1{|i-j|},\quad b=\frac1{N+1-i-j}.
\end{equation}
Here $i,j$ are the left members of the pairs, so $a>b>0$. The second term flips the swap parity of both pairs and has a nonzero block for each of the four choices of initial pair parities. For an explicit check, write $|s\rangle=(|\uparrow\downarrow\rangle-
|\downarrow\uparrow\rangle)/\sqrt2$ and
$|t_\pm\rangle=|\uparrow\uparrow\rangle,|\downarrow\downarrow\rangle$.
For $K=A_i^+A_j^-+A_i^-A_j^+$,
\begin{align}
 K|s,s\rangle&=-2(|t_+,t_-\rangle+|t_-,t_+\rangle),\\
 K|s,t_+\rangle&=2|t_+,s\rangle,\qquad
 K|t_+,t_-\rangle=-2|s,s\rangle.
\end{align}
Hermitian conjugation or pair interchange gives the fourth parity block.
These nonzero matrix elements establish the claim without assuming an
invertible transition between the differently sized sectors. No other pair interaction contributes to that same change of precisely these two parity labels.

Commutation with the exchange Hamiltonian gives $x_\varepsilon=x_{\varepsilon'}$ whenever $\varepsilon'$ is obtained by flipping two labels. Such moves connect all sign strings with a fixed product $\prod_i\varepsilon_i$. Thus only two scalars remain, giving $X=\alpha\fnIdentity+\beta\prod_iF_i$. For a single pair each product class contains one string and the same conclusion holds.
\end{fnproof}

\subsection{Odd chains}
Let $N=2m+1$ and $c=m+1$. Lemma~\ref{fn:alg:lem:peel} supplies $X_c,Y_c,Z_c\in\fnAlg$. Conjugating $D$ by $X_c$ isolates
\begin{equation}
 D_{c,\mathrm{rest}}=\tfrac12(D-X_cDX_c)
 =Z_c V,\qquad V=\sum_{i\ne c}\frac{Z_i}{|i-c|}.
\end{equation}
Multiplication by $Z_c$ shows $V\in\fnAlg$. The rest collective spin also belongs to $\fnAlg$. In particular,
\begin{equation}
 (\fnAd_V)^k S_{\mathrm{rest}}^+
 =2^k\sum_{i\ne c}|i-c|^{-k}\sigma_i^+.
\end{equation}
The $m$ distinct values $1,1/2,\ldots,1/m$ distinguish all reflection pairs. Vandermonde interpolation, using $k=0,\ldots,m-1$, yields $S_{\{i,r_i\}}^+\in\fnAlg$ for every pair, and hence all three collective components of each pair.

Any $X\in\fnAlg'$ is the identity on the central qubit tensored with an operator $X_R$ on the rest, because it commutes with the full matrix algebra of that qubit. It commutes with every independent pair rotation. Taking the partial trace over the central qubit in $[X,B_N]=0$ gives $[X_R,B_R]=0$, since cross terms have zero central trace. Lemma~\ref{fn:alg:lem:pairs} implies
$X_R=\alpha\fnIdentity+\beta\prod_iF_i$. Reflection fixes the central site, so this proves Eq.~\eqref{fn:alg:eq:goal} for odd $N\geq3$.

\subsection{Even chains: triplet block and elimination of a dark singlet block}
The case $N=2$ follows directly from the spin-one plus spin-zero decomposition. Let $N=2m\geq4$, and let $a=m$, $b=m+1$ be the central pair. Its collective Pauli operators $S_0^\alpha=\sigma_a^\alpha+\sigma_b^\alpha$ belong to $\fnAlg$ by Lemma~\ref{fn:alg:lem:peel}. So do its triplet and singlet projectors,
\begin{equation}
 P_t=\frac18\sum_{\alpha=x,y,z}(S_0^\alpha)^2,\qquad P_s=\fnIdentity-P_t.
\end{equation}
Operators on the central pair are implicitly embedded with the identity on the remaining sites $R$. In explicit tensor-product decompositions below, $P_s$ and $\fnIdentity_{s,t}$ denote the unembedded central-pair operators.

\subsubsection{Resolving every other pair within the central triplet}
A collective $\pi$ rotation about $x$ on the central pair isolates from $D$ the interaction between that pair and $R$:
\begin{equation}
 D_{0R}=Z_a\sum_{i\in R}c_{ai}Z_i+Z_b\sum_{i\in R}c_{bi}Z_i,
 \qquad c_{ji}=|j-i|^{-1}.
\end{equation}
Since $P_t(Z_a-Z_b)P_t=0$,
\begin{equation}
 P_tD_{0R}P_t=\frac12S_0^z V_+,
 \quad V_+=\sum_{i\in R}(c_{ai}+c_{bi})Z_i.
\end{equation}
Central collective rotations also give $S_0^xV_+/2$ and $S_0^yV_+/2$. Multiplying by $S_0^\alpha$ and summing yields $4P_tV_+\in\fnAlg$. Also $P_tS_R^+\in\fnAlg$. Repeated commutators thus produce
\begin{equation}
 P_t\sum_{i\in R}(c_{ai}+c_{bi})^k\sigma_i^+\in\fnAlg.
\end{equation}
For a reflection pair at distance $r$ from the nearer central site, the common weight is $v_r=1/r+1/(r+1)$, $r=1,\ldots,m-1$. These positive weights are strictly decreasing. Vandermonde interpolation resolves every remaining reflection pair conditionally on the central triplet:
\begin{equation}
 P_tS_{\{i,r_i\}}^{x,y,z}\in\fnAlg.
 \label{fn:alg:eq:conditionalcontrols}
\end{equation}

For $X\in\fnAlg'$, central Schur decomposition gives
\begin{equation}
 X=\fnIdentity_t\otimes X_t+\fnIdentity_s\otimes X_s.
\end{equation}
Equation~\eqref{fn:alg:eq:conditionalcontrols} implies that $X_t$ commutes with all pair rotations on $R$. Taking the trace over the three-dimensional central triplet in the triplet block of $[X,B_N]=0$ gives $[X_t,B_R]=0$: the projected cross interaction is linear in the traceless spin-one generators, and the internal central exchange contributes only a scalar to this partial trace. By Lemma~\ref{fn:alg:lem:pairs},
\begin{equation}
 X_t=\alpha\fnIdentity_R+\beta\Pi_R,\qquad \Pi_R=\prod_{i\in R\text{ pairs}}F_i.
\end{equation}
Full reflection is $F_0\Pi_R$, where $F_0$ is $+1$ on the central triplet and $-1$ on its singlet. Consequently
\begin{equation}
 Y=X-\alpha\fnIdentity-\beta\Pi_N=P_s\otimes Y_s
 \label{fn:alg:eq:remainder}
\end{equation}
for an operator $Y_s$ on $R$. It remains to prove $Y_s=0$.

\subsubsection{The remaining singlet block cannot survive}
Because $Y$ commutes with global spin, $[Y_s,J_R^\pm]=0$. The singlet-to-triplet block of $[Y,B_N]=0$ gives
\begin{equation}
 W_+Y_s=W_-Y_s=0,\qquad
 W_\pm=\sum_{i\in R}w_i\sigma_i^\pm,\quad w_i=c_{ai}-c_{bi}.
 \label{fn:alg:eq:dark}
\end{equation}
Indeed, acting on the central singlet, the cross exchange produces the polarized triplets with rest operators proportional to $W_-$ and $W_+$, respectively; their central states are orthogonal.

Set $W_z=\sum_{i\in R}w_i Z_i=[W_+,J_R^-]$. From Eq.~\eqref{fn:alg:eq:dark} and $[Y_s,J_R^-]=0$ it follows that
\begin{equation}
 W_zY_s=0.
\end{equation}
If two operators annihilate $Y_s$ by left multiplication, their commutator does as well. Repeated commutation of $W_\pm$ with $W_z$ therefore implies
\begin{equation}
 \left(\sum_{i\in R}w_i^k\sigma_i^\pm\right)Y_s=0
 \qquad(k=1,2,\ldots).
\end{equation}
The $N-2$ numbers $w_i$ are precisely
\begin{equation}
 \left\{\frac1{r(r+1)},-\frac1{r(r+1)}:
 r=1,\ldots,m-1\right\}.
\end{equation}
They are all nonzero and pairwise distinct. The matrix $(w_i^k)_{k=1}^{N-2}$ is an invertible Vandermonde matrix times a nonzero diagonal matrix. It follows that
\begin{equation}
 \sigma_i^+Y_s=\sigma_i^-Y_s=0\qquad\text{for every }i\in R.
\end{equation}
For any one such site, $\sigma_i^-\sigma_i^++\sigma_i^+\sigma_i^-=\fnIdentity$ forces $Y_s=0$. This closes the even case. Finally, both $\fnIdentity$ and $\Pi_N$ commute with the generators by direct inspection, proving equality in Eq.~\eqref{fn:alg:eq:goal}.

\paragraph{Number-theoretic input.}
The only external theorem used in the construction is Bertrand's theorem:
for every integer $n>1$ there is a prime strictly between $n$ and $2n$.
An accessible proof is given in Ref.~\cite{Galvin2015}.

\section{Finite-size spectral reduction and strong-Mpemba existence}
\label{sec:fn_dyn}
\subsection{Model and precise quantifiers}
We work with the interaction-picture generator in
Eqs.~\eqref{fn:dyn:eq:model}--\eqref{eq:Hbar}, the stationary state in
Eq.~\eqref{fn:dyn:eq:thermal}, and the family in Eq.~\eqref{eq:family}.
Throughout this proof $N\geq2$ is fixed and finite, $E>0$ is fixed,
and $d\to0^+$. All decay rates in this section refer to the reduced time
$s=g^2\eta$; thus $\gamma_s$ here equals $\bar\gamma_{\rm slow}$ in the
Letter. A common local shift has been removed as in Sec.~\ref{sec:model}.

Let $\Pi$ reverse the site order, $P_\pm=(\fnIdentity\pm\Pi)/2$, and
\begin{equation}
 t=\fnTr(\Pi\rho_\beta)=\left[\frac{1+q^2}{(1+q)^2}\right]^{\lfloor N/2\rfloor},
 \qquad w_\pm=\frac{1\pm t}{2},\qquad \rho_\pm=\frac{P_\pm\rho_\beta}{w_\pm}.
\end{equation}
In particular, $0<t<1$.

\begin{fntheorem}[Finite-$N$ strong Mpemba existence]\label{fn:dyn:thm:full}
For every finite $N\geq2$ and every $E>0$, there exists $d_0(N,E)>0$ such that for each $0<d<d_0$:
\begin{enumerate}
 \item $\rho_\beta$ is the unique stationary state. The generator has a simple real slow eigenvalue $-\gamma_s(d)$, with
 \begin{equation}
 \gamma_s(d)=\kappa_{N,E}d^2+O(d^4),\qquad
 \kappa_{N,E}=\frac{N(N^2-1)(E+E^{-1})}{36\pi\sinh(2\pi E)(1-t^2)}>0.
 \label{fn:dyn:eq:kappa}
 \end{equation}
 All other nonstationary eigenvalues have real part at most $-\delta_{N,E}$ for a positive constant independent of sufficiently small $d$.
 \item There is a unique $p_\star(d)\in(0,1)$ within this affine preparation family that cancels the slow mode exactly:
 \begin{equation}
 p_\star(d)=t+O(d^2),\qquad
 \fnP_s(d)[\rho_A(p_\star)-\rho_\beta]=0.
 \end{equation}
 \item A physical, explicitly specified reference state $\rho_B$, independent of $d$, obeys
 \begin{equation}
 D_A(0)>D_B(0)>0,\qquad
 \lim_{s\to\infty}\frac{D_A(s)}{D_B(s)}=0,
 \quad D_X(s)=\tfrac12\fnNorm{e^{s\fnL_d}\rho_X-\rho_\beta}_1.
 \label{fn:dyn:eq:crossing}
 \end{equation}
 Hence at least one crossing occurs at a finite positive time, and $A$ remains closer to equilibrium than $B$ for all sufficiently late times.
\end{enumerate}
The constants and the allowed distance may depend on $N,E$. No uniform-in-$N$ interval is asserted.
\end{fntheorem}

\subsection{From the commutant to a controlled fast complement}
At $d=0$ in the collective dissipative terms, the Hilbert-space reflection
$\Pi$ commutes separately with the Hamiltonian and with the collective
jumps $J_\pm$: this is a strong symmetry that conserves populations of
$P_+$ and $P_-$. For finite $d$, each spatial reflection eigenchannel has
a definite sign under conjugation by $\Pi$; the corresponding dissipator
is reflection covariant even when an odd channel transfers population
between those sectors. Thus the full weak symmetry of the Lindbladian
does not protect the sector populations. In particular the slow left
observable is even under reflection conjugation, so its cancellation is
not the removal of a reflection-odd weak-symmetry eigenmode.

Theorem~\ref{fn:alg:thm:main} above establishes
\begin{equation}
 \{H_{-1},J_+,J_-\}'=\operatorname{span}\{\fnIdentity,\Pi\},\qquad
 H_{-1}=-\frac{E}{4\pi}\sum_{i\ne j}\frac{\sigma_i^+\sigma_j^-}{|i-j|},
 \quad J_\pm=\sum_i\sigma_i^\pm.
 \label{fn:dyn:eq:commutant}
\end{equation}
It covers both even and odd sizes without a numerical-rank assumption. The positive value $E>0$ ensures the prefactor is nonzero.

Split the generator exactly as
\begin{equation}
 \fnL_d=\fnL_0(d)+V(d),\quad
 \fnL_0(d)=-i[\bar H(d),\cdot]+a_\downarrow\fnD[J_-]+a_\uparrow\fnD[J_+].
\end{equation}
All operator-norm estimates below are at fixed finite $N,E$. The perturbation $V$ need not itself be a GKLS generator. We have $\bar H(d)=H_{-1}/d+O(d)$ and
$V(d)=d^2V_2+O(d^4)$. The symmetry-sector structure is consistent with the general theory of strong symmetries~\cite{buvca2012note}. Both sector states $\rho_\pm$ are stationary for $\fnL_0(d)$ for every positive $d$. Define
\begin{equation}
 \fnP X=\rho_+\fnTr(P_+X)+\rho_-\fnTr(P_-X),\qquad \fnQ=1-\fnP.
\end{equation}
Then $\fnL_0\fnP=\fnP\fnL_0=0$. In the Heisenberg picture, $\fnP^\dagger$ is the orthogonal projection onto $\operatorname{span}\{\fnIdentity,\Pi\}$ for
$\langle X,Y\rangle_\beta=\fnTr(\rho_\beta X^\dagger Y)$.

\paragraph{Bounded reduced inverse.}
The GKLS dissipation identity and stationarity give
\begin{equation}
 -2\operatorname{Re}\langle X,\fnL_0^\dagger X\rangle_\beta
 =a_\downarrow\fnNorm{[J_-,X]}_\beta^2+a_\uparrow\fnNorm{[J_+,X]}_\beta^2.
 \label{fn:dyn:eq:dissipation}
\end{equation}
For completeness, Eq.~\eqref{fn:dyn:eq:dissipation} follows by averaging
\begin{equation}
 \fnL_0^\dagger(X^\dagger X)-(\fnL_0^\dagger X^\dagger)X
 -X^\dagger\fnL_0^\dagger X
 =\sum_\alpha[L_\alpha,X]^\dagger[L_\alpha,X]
\end{equation}
in the stationary state, with $L_\alpha=\sqrt{a_\downarrow}J_-$ and
$\sqrt{a_\uparrow}J_+$. It applies also to non-Hermitian $X$ and requires
stationarity and faithfulness, not normality of the superoperator.

If $\fnL_0^\dagger$ had no uniform lower singular-value bound on $\operatorname{ran}\fnQ^\dagger$ as $d\to0$, there would be normalized $X_n$ in that space and $d_n\to0$ with $\fnL_0^\dagger(d_n)X_n\to0$. Finite-dimensional compactness gives a unit-norm limit $X_*$. Equation~\eqref{fn:dyn:eq:dissipation} implies $[J_\pm,X_*]=0$. Multiplying the generator equation by $d_n$ also gives $[H_{-1},X_*]=0$. Equation~\eqref{fn:dyn:eq:commutant} puts $X_*$ in $\operatorname{ran}\fnP^\dagger$, a contradiction. Thus the reduced inverse
\begin{equation}
 S(d)=(\fnQ\fnL_0(d)\fnQ)^{-1}\big|_{\operatorname{ran}\fnQ}
\end{equation}
is uniformly bounded for sufficiently small $d$, at fixed $N,E$. This also establishes that the reference kernel is exactly two-dimensional.

\paragraph{Excluding nearly undamped oscillatory modes.}
An inverse bound near zero alone does not exclude eigenvalues near the
imaginary axis at diverging imaginary frequency. We prove the following
limiting observation for use both here and for the full generator.
Let $d_n\to0$ and let normalized ordinary eigenobservables satisfy
\begin{equation}
 \mathcal A_nX_n=\lambda_nX_n,\quad \fnNorm{X_n}_\beta=1,
 \quad\operatorname{Re}\lambda_n\to0,
 \qquad \mathcal A_n=\fnL_0^\dagger(d_n)+W_n,
 \quad\fnNorm{W_n}=O(d_n^2).
\end{equation}
We will use $W_n=0$ or $V(d_n)^\dagger$. Every eigenvalue has an ordinary
eigenvector, even in a defective Jordan block. No continuous choice of
eigenvectors, simplicity, or diagonalizability is required. Compactness
in this fixed finite-dimensional space gives a subsequence
$X_n\to X_*$ with $\fnNorm{X_*}_\beta=1$. Equation~\eqref{fn:dyn:eq:dissipation} gives
\begin{equation}
 a_\downarrow\fnNorm{[J_-,X_n]}_\beta^2+
 a_\uparrow\fnNorm{[J_+,X_n]}_\beta^2
 =-2\operatorname{Re}\lambda_n
   +2\operatorname{Re}\langle X_n,W_nX_n\rangle_\beta\longrightarrow0.
\end{equation}
Both coefficients are strictly positive, hence $[J_\pm,X_*]=0$.
Furthermore $d_n\mathcal A_n$ is bounded and therefore
$|d_n\lambda_n|\leq\fnNorm{d_n\mathcal A_n}$ is bounded. Passing to another
subsequence yields $d_n\lambda_n\to i\omega$ with $\omega\in\mathbb R$.
Multiplying the eigenvalue equation by $d_n$ yields
\begin{equation}
 [H_{-1},X_*]=\omega X_*.
 \label{fn:dyn:eq:limitfrequency}
\end{equation}
This step explicitly allows $|\operatorname{Im}\lambda_n|\to\infty$.

For any nonzero block $Y=P_aX_*P_b$, $a,b\in\{+,-\}$, reflection commutes
with $H_{-1},J_\pm$, so $[J_\pm,Y]=0$ and $[H_{-1},Y]=\omega Y$.
Taking adjoints gives $[H_{-1},Y^\dagger]=-\omega Y^\dagger$; consequently
both $Y^\dagger Y$ and $YY^\dagger$ commute with all three generators.
Equation~\eqref{fn:dyn:eq:commutant} and their supports imply
\begin{equation}
 Y^\dagger Y=c_bP_b,\qquad YY^\dagger=c_aP_a,
 \qquad c_a,c_b>0.
\end{equation}
Thus $Y$ is both injective and surjective between the two sectors.
Their dimensions are
\begin{equation}
 D_\pm=\frac{2^N\pm2^{\lceil N/2\rceil}}2,
\end{equation}
since the trace of reflection counts the $2^{\lceil N/2\rceil}$ palindromic
basis strings. These dimensions differ, so $a=b$. On this sector $Y$
is invertible, and Eq.~\eqref{fn:dyn:eq:limitfrequency} gives
$H_a-YH_aY^{-1}=\omega\fnIdentity_a$. Its trace forces $\omega D_a=0$.
Thus $\omega=0$, and the commutant theorem finally gives
\begin{equation}
 X_*\in\operatorname{span}\{\fnIdentity,\Pi\}=\operatorname{ran}\fnP^\dagger.
 \label{fn:dyn:eq:limitinP}
\end{equation}
For the reference complement, $\fnP^\dagger X_n=0$ and hence
$\fnP^\dagger X_*=0$, contradicting Eq.~\eqref{fn:dyn:eq:limitinP} and unit norm.
Therefore a uniform positive damping gap exists on that complement.

\subsection{The slow eigenvalue and its projector}
Write $A_Q=\fnQ\fnL_0\fnQ|_{\operatorname{ran}\fnQ}$ and choose a constant
$C$ with $\fnNorm{A_Q^{-1}}\leq C$ for all sufficiently small $d$. For
$|z|<1/(2C)$,
\begin{equation}
 (A_Q-z)^{-1}=(\fnIdentity-zA_Q^{-1})^{-1}A_Q^{-1},\qquad
 \fnNorm{(A_Q-z)^{-1}}\leq2C.
\end{equation}
On a fixed circle $|z|=\eta<1/(2C)$, the full reference resolvent is uniformly
bounded too: the stationary block contributes $z^{-1}\fnP$ and the complementary
block has the bound just given. The resolvent identity and $V=O(d^2)$ then
keep this circle in the resolvent set throughout $\fnL_0+\theta V$,
$0\leq\theta\leq1$. Define its Riesz projection by the positively oriented contour
\begin{equation}
 \fnP_{\rm cl}(d)=\frac{1}{2\pi i}\oint_{|z|=\eta}
 (z-\fnL_d)^{-1}\,dz.
 \label{fn:dyn:eq:rieszcluster}
\end{equation}
The homotopy preserves its rank, which is two. The resolvent identity gives
$\fnP_{\rm cl}(d)=\fnP+O(d^2)$ in operator norm. This rank counts algebraic
multiplicity and does not assume that the enclosed eigenvalues are semisimple. Zero is semisimple for the reference generator because its
complement is invertible and $\fnL_0\fnP=\fnP\fnL_0=0$.

The scale $\lambda=O(d^2)$ must be derived rather than assumed. For an
eigenvector in this disk write $x=u+v$, with $u=\fnP x$ and $v=\fnQ x$.
The complementary resolvent remains bounded after adding $\fnQ V\fnQ$, so
\begin{equation}
 v=-(A_Q+\fnQ V\fnQ-\lambda)^{-1}\fnQ V u,
 \qquad \fnNorm v\leq C' d^2\fnNorm u.
\end{equation}
In particular $u\ne0$. The projected equation
$\lambda u=\fnP V(u+v)$ gives $|\lambda|\leq C''d^2$.
Eliminating $v$ therefore gives
\begin{equation}
 W_{\rm eff}(\lambda,d)
 =\fnP V\fnP-\fnP V\fnQ(\fnQ\fnL_0\fnQ+\fnQ V\fnQ-\lambda)^{-1}\fnQ V\fnP
 =\fnP V\fnP+O(d^4).
 \label{fn:dyn:eq:schur}
\end{equation}
This estimate retains the singular exchange Hamiltonian inside $\fnL_0$; it never substitutes $d=0$ into that Hamiltonian.

Let $R$ now denote site reflection, $R_{ij}=\delta_{i,N+1-j}$, and set
\begin{equation}
 F_o=\frac{I_N-R}{2}F\frac{I_N-R}{2},\quad
 T_o=\fnTr F_o=\frac12\left(N-\sum_i F_{i,N+1-i}\right).
\end{equation}
Only odd spatial channels transfer population between parity sectors. For a normalized odd channel $A=\sum_i v_i\sigma_i^-$, the thermal expectations of $A^\dagger A$ and $AA^\dagger$ are $u$ and $1-u$. The parity-weighted expectation of
$a_\downarrow A^\dagger A+a_\uparrow AA^\dagger$ is zero, by $\Pi A\Pi=-A$ and the thermal relation for $A$. It follows that the matrix of $\fnP V\fnP$ on $(\rho_+,\rho_-)$ is
\begin{equation}
 W=\begin{pmatrix}-k_+&k_-\\ k_+&-k_-\end{pmatrix},\qquad
 k_\pm=\frac{2a_\downarrow uT_o}{1\pm t}.
\end{equation}
These are exact expressions for this projected matrix, not the exact rates of the full generator.

Writing $x_i=i-(N+1)/2$ and $b(E)=(1+E^{-2})/6$,
\begin{equation}
 F_{ij}=1-b(E)(i-j)^2d^2+O(d^4),\quad
 F_o=2b(E)d^2xx^{\mathsf T}+O(d^4),\quad
 \sum_i x_i^2=\frac{N(N^2-1)}{12}.
\end{equation}
Thus $k_++k_-=\kappa_{N,E}d^2+O(d^4)$ with Eq.~\eqref{fn:dyn:eq:kappa}. To justify the splitting despite the eigenvalue-dependent Schur complement,
write $\lambda=d^2\mu$ and let $W_2$ be the matrix of $\fnP V_2\fnP$ in the
sector basis. The reduced characteristic equation is
\begin{equation}
 0=\det\big[\mu I_2-W_2+O(d^2)\big],\qquad
 \det(\mu I_2-W_2)=\mu(\mu+\kappa_{N,E}).
\end{equation}
The error is uniform on bounded $\mu$ contours, where the complementary
resolvent is analytic. Small disjoint contours around $0$ and
$-\kappa_{N,E}$ therefore contain one root each, counted with multiplicity,
by the argument principle (or Rouch\'e's theorem). The roots are simple and
shift by $O(d^2)$ in $\mu$, because the limiting roots are simple. The first
root is exactly zero by stationarity of $\rho_\beta$. The other is thus
$\lambda_s=-\kappa_{N,E}d^2+O(d^4)$. To establish reality, use Hermiticity preservation:
\begin{equation}
 \fnL_d(X^\dagger)=(\fnL_dX)^\dagger.
\end{equation}
Every nonreal eigenvalue therefore has its distinct complex conjugate
as another eigenvalue. The fixed disk in Eq.~\eqref{fn:dyn:eq:rieszcluster} is
invariant under complex conjugation and contains exactly two roots counted
with algebraic multiplicity. One is the simple stationary root zero.
If the other root $\lambda_s$ were nonreal, $0,\lambda_s,\lambda_s^*$
would give at least three roots in that disk, a contradiction.
Hence $\lambda_s$ is real and, by $\kappa_{N,E}>0$, strictly negative
for sufficiently small $d$. No time-reversal symmetry, detailed balance
of the full generator, or diagonalizability of its fast block is used.

\paragraph{Uniform gap for the full fast spectral complement.}
Suppose no positive gap bounded away from zero existed outside the cluster
for all sufficiently small $d$. Since $\fnL_d$ is a finite-dimensional GKLS
generator its eigenvalues have nonpositive real parts (its bounded
semigroup excludes exponential growth). We could therefore choose
$d_n\to0$ and eigenvalues $\lambda_n$ of $\fnL_{d_n}^\dagger$ outside the
conjugate cluster with $\operatorname{Re}\lambda_n\to0$.
Choose normalized ordinary eigenobservables $X_n$ for them. The preceding
limiting argument applies with $W_n=V(d_n)^\dagger$ and proves
Eq.~\eqref{fn:dyn:eq:limitinP}. On the other hand, spectral projection gives
\begin{equation}
 \fnP_{\rm cl}(d_n)^\dagger X_n=0.
\end{equation}
Indeed, the Riesz projection annihilates every eigenspace, and even every
generalized eigenspace, associated with eigenvalues outside its contour.
Since $\fnP_{\rm cl}(d_n)^\dagger\to\fnP^\dagger$, taking the limit gives
$\fnP^\dagger X_*=0$. Together with Eq.~\eqref{fn:dyn:eq:limitinP}, this forces
$X_*=0$, contradicting $\fnNorm{X_*}_\beta=1$. Thus there exist
$\delta_{N,E}>0$ and a sufficiently small distance interval on which every
nonstationary eigenvalue other than $\lambda_s$ has real part at most
$-\delta_{N,E}$. This argument includes coalescing eigenvalues and
arbitrarily large imaginary parts; no uniform eigenvector condition number
is asserted or needed.

Normalize the Hermitian slow left and right eigenoperators by
\begin{equation}
 \ell_s=\frac{\Pi-t\fnIdentity}{2}+O(d^2),\qquad
 R_s=\rho_+-\rho_-+O(d^2),\qquad
 \fnTr(\ell_sR_s)=1.
\end{equation}
These estimates can also be seen without assuming diagonalizability
of the fast block. A nonzero right eigenvector is traceless, so its nonzero
$\fnP$ component is proportional to $\rho_+-\rho_-$. The bounded complementary
resolvent makes its $\fnQ$ component $O(d^2)$ relative to that component.
Similarly, $\fnTr(\ell_s\rho_\beta)=0$ forces the nonzero $\fnP^\dagger$ component
of the left eigenoperator to be proportional to $\Pi-t\fnIdentity$, and its
$\fnQ^\dagger$ component is $O(d^2)$. The leading pairing equals one, permitting
the displayed biorthogonal normalization. The simple real eigenvalue allows
both eigenoperators to be chosen Hermitian. Also $\fnTr(\ell_s\rho_\beta)=0$ exactly. Thus
\begin{equation}
 \fnP_s X=R_s\fnTr(\ell_s X).
\end{equation}

The reflection symmetry has distinct roles in the reference
and full dynamics. For $\mathcal L_0(d)$, reflection is a strong
symmetry: $\Pi$ commutes with the Hamiltonian and with each
collective jump operator $J_\pm$, so the populations
$\operatorname{Tr}(P_\pm\rho)$ are conserved. At finite separation,
the full generator instead retains the conjugation symmetry
\begin{equation}
    \mathscr R(X)=\Pi X\Pi,
    \qquad
    \mathcal L_d\mathscr R=\mathscr R\mathcal L_d.
    \label{eq:SM_reflection_conjugation}
\end{equation}
Indeed, the spatial rate matrix is invariant under site reversal,
so its jump channels can be chosen with definite reflection parity.
Odd channels transfer population between the Hilbert-space
reflection sectors, as quantified by the projected transfer
matrix derived above, while their dissipators preserve
the conjugation symmetry.

Since the slow eigenvalue is simple, its left eigenspace is
invariant under $\mathscr R$. Hence
$\Pi\ell_s\Pi=\pm\ell_s$. The leading form
$\ell_s=(\Pi-tI)/2+O(d^2)$ derived above is reflection-even and nonzero; continuity therefore fixes
\begin{equation}
    \Pi\ell_s\Pi=\ell_s
    \label{eq:SM_slow_observable_even}
\end{equation}
for sufficiently small positive $d$.
Thus, reflection invariance of an initial density matrix does
not by itself eliminate this slow mode. Its cancellation requires
$\operatorname{Tr}[\ell_s(\rho-\rho_\beta)]=0$, which reduces
at leading order to matching the equilibrium reflection-sector
weights, $\operatorname{Tr}(\Pi\rho)=t$.

\subsection{Exact cancellation in the proposed preparation family}
Put $c_j(d)=\fnTr[\ell_s(\rho_j-\rho_\beta)]$. The parity expectations are $\fnTr(\Pi\rho_0)=0$ and $\fnTr(\Pi\rho_1)=1$, giving
\begin{equation}
 c_0=-\frac t2+O(d^2),\qquad
 c_1=\frac{1-t}{2}+O(d^2).
\end{equation}
For sufficiently small $d$, $c_0<0<c_1$. The exact unique root is
\begin{equation}
 \boxed{p_\star(d)=-\frac{c_0(d)}{c_1(d)-c_0(d)}=t+O(d^2)\in(0,1).}
 \label{fn:dyn:eq:pstar}
\end{equation}
It follows directly that $\fnP_s[\rho_A(p_\star)-\rho_\beta]=0$. No claim of a two-mode decomposition of the entire state is made: many faster modes can remain.

\subsection{A constructive reference state and a rigorous crossing}
Define the equilibrium probabilities of the two basis states used in $A$:
\begin{equation}
 v_0=\frac{q}{(1+q)^N},\qquad v_1=\frac{q^N}{(1+q)^N},\qquad
 r=1-v_0-v_1>0.
\end{equation}
Here $r>0$ because $\rho_\beta$ has strictly positive weight on basis states outside this two-dimensional support. Set
\begin{equation}
 \boxed{\varepsilon=\frac{r}{2(1-v_1)},\qquad
 \rho_B=(1-\varepsilon)\rho_\beta+\varepsilon\rho_1.}
 \label{fn:dyn:eq:B}
\end{equation}
Since $0<r<1-v_1$, $0<\varepsilon<1/2$ and $\rho_B$ is a faithful physical state independent of $d$.

Let $P_A$ project onto the span of the two preparation kets. Measuring $P_A$ gives the trace-distance lower bound
\begin{equation}
 D_A(0)\geq\big|\fnTr[P_A(\rho_A-\rho_\beta)]\big|=r.
\end{equation}
Because $\rho_1$ and $\rho_\beta$ are diagonal in the same basis,
\begin{equation}
 D_B(0)=\varepsilon D(\rho_1,\rho_\beta)
 =\varepsilon(1-v_1)=\frac r2<D_A(0).
 \label{fn:dyn:eq:initialordering}
\end{equation}
Meanwhile, the reference state's slow coefficient is
\begin{equation}
 c_B=\fnTr[\ell_s(\rho_B-\rho_\beta)]=\varepsilon c_1>0.
\end{equation}
The unique stationary spectral projector is
$\fnP_{\rm ss}X=\rho_\beta\fnTr X$, since the stationary eigenvalue is simple
and the generator preserves trace. It annihilates both initial deviations.
After also removing the slow projection of $A$, finite-dimensional spectral
decomposition at each fixed sufficiently small $d$ gives
\begin{align}
 \rho_A(s)-\rho_\beta&=O(e^{-\delta s/2}),\\
 \rho_B(s)-\rho_\beta&=c_B e^{-\gamma_s s}R_s+O(e^{-\delta s/2}).
\end{align}
To make the role of Jordan blocks explicit, a block $\lambda I+K$ of size
$m$, with $K^m=0$, contributes
\begin{equation}
 e^{s(\lambda I+K)}=e^{\lambda s}\sum_{j=0}^{m-1}\frac{s^jK^j}{j!}.
\end{equation}
For each fixed $d$, $s^j e^{-\delta s/2}$ is bounded on $s\geq0$.
Thus these polynomial factors and the finite change-of-basis factors can
be absorbed into a constant multiplying $e^{-\delta s/2}$. The resulting
constant may depend on $d,N,E$ and is not claimed uniform as $d\to0$.
Only the spectral gap $\delta_{N,E}$ is uniform in sufficiently small $d$.
This fixed-$d$ estimate suffices for the following asymptotic crossing
argument; no diagonalizability assumption is needed. Shrink $d_0$ so that $\gamma_s<\delta/2$. Then
\begin{equation}
 D_B(s)=\frac{c_B\fnNorm{R_s}_1}{2}e^{-\gamma_s s}[1+o(1)],\qquad
 \frac{D_A(s)}{D_B(s)}\longrightarrow0.
\end{equation}
Together with Eq.~\eqref{fn:dyn:eq:initialordering}, continuity proves a finite positive crossing and eventual permanent reversal of the initial ordering. This completes Theorem~\ref{fn:dyn:thm:full}.

\subsection{Interpretation and remaining boundaries}
The conclusion is an existence theorem for a strong quantum Mpemba effect ordered by trace distance, for every finite detector number in this model. The isolated slow sector also provides the spectral separation associated with metastable relaxation~\cite{macieszczak2016towards}. The spectral cancellation is exact at $p_\star(d)$; $t$ is only its leading approximation. The reference state in Eq.~\eqref{fn:dyn:eq:B} is constructed for the theorem and is not asserted to coincide with the numerical reference state in Eq.~\eqref{eq:Bnum}.

The theorem does not order Gibbs temperatures of the preparations, prove energy-distance crossings, establish uniqueness of the trace-distance crossing, or give a useful numerical upper bound on its time. It supplies no distance interval uniform in $N$, no thermodynamic limit at fixed spacing, and no microscopic finite-coupling or finite-profile error control. In particular it does not certify $d=0.01$ for arbitrary $N$. Within the stated pointlike GKLS model, however, it closes both the general-size algebraic premise and the existence of a relaxation-order reversal for the explicit pair above.

\subsection{Common thermal-field corollary}\label{sec:thermal_corollary}
The same constructive result holds for the common thermal field of
Sec.~\ref{sec:thermal_corollary_benchmark}, not only for the accelerated
vacuum. At fixed finite $N\geq2$ and $E>0$, the inertial thermal generator
has the same local coefficients $a_\downarrow,a_\uparrow$, stationary
state $\rho_\beta$, and small-$d$ leading exchange Hamiltonian as the
accelerated generator:
\begin{equation}
 \bar J_{ij}^{\rm th}
 =-\frac{E}{4\pi|i-j|d}+O(d),\qquad i\ne j.
\end{equation}
The thermal spatial factor has the expansion
\begin{equation}
 F^{\rm th}_{ij}=\frac{\sin[(i-j)d]}{(i-j)d}
 =1-\frac{(i-j)^2d^2}{6}+O(d^4),
 \qquad F^{\rm th}_{ii}=1.
 \label{fn:dyn:eq:thermal_expansion}
\end{equation}
Its rate matrix is positive semidefinite at every separation, since
\begin{equation}
 F^{\rm th}_{ij}=\frac12\int_{-1}^{1}
 e^{i(i-j)d z}\,dz
\end{equation}
is a Gram matrix. The shared inverse-distance exchange and collective
jumps give the same reflection commutant and fast-complement gap argument.
The Schur reduction of Theorem~\ref{fn:dyn:thm:full} therefore carries over
with $b_{\rm th}=1/6$ in the expansion of $F^{\rm th}$, rather than
$b_U=(1+E^{-2})/6$. More explicitly, the reflection-odd rate matrix is
$F_o^{\rm th}=2b_{\rm th}d^2\mathbf x\mathbf x^{\mathsf T}+O(d^4)$,
where $x_i=i-(N+1)/2$, so the effective sector-transfer rates yield
\begin{equation}
 \boxed{\gamma_s^{\rm th}(d)=\kappa_{N,E}^{\rm th}d^2+O(d^4),\qquad
 \kappa_{N,E}^{\rm th}
 =\frac{N(N^2-1)E}{36\pi\sinh(2\pi E)(1-t^2)}.}
 \label{fn:dyn:eq:thermal_kappa}
\end{equation}
In particular, this thermal generator has, for each sufficiently small
positive $d$, a unique stationary state and a simple isolated slow mode.
The same affine family has a unique physical preparation
$p_\star^{\rm th}(d)=t+O(d^2)$ with an exactly vanishing slow spectral
projection. The reference state in Eq.~\eqref{fn:dyn:eq:B} retains a
nonzero slow component and produces a trace-distance crossing and an
eventual reversal of relaxation ordering. Thus the thermal comparison
extends to every fixed finite $N$, not merely the numerical $N=4$ case.
Combining Eqs.~\eqref{fn:dyn:eq:kappa} and
\eqref{fn:dyn:eq:thermal_kappa} yields
\begin{equation}
 \boxed{\lim_{d\to0^+}\frac{\gamma_s^{U}(d)}
 {\gamma_s^{\rm th}(d)}=1+E^{-2}=1+\frac{a^2}{\omega_0^2}.}
 \label{fn:dyn:eq:thermal_ratio}
\end{equation}
Here $a$ fixes the matching thermal temperature; the ratio concerns the
slow sector-relaxation rates, not the enhancement factors obtained by
removing those modes. This is a fixed-$N$, fixed-$E$, small-$d$ result
within the respective pointlike secular generators.

\subsection{Explicit Gram representation of the rate matrix}\label{fn:dyn:app:gram}
This subsection closes the positivity premise directly for the displayed kernel,
without relying on its interpretation as a field response. Set
$\mathbf y_i=(id/E,0)\in\mathbb R^2$ and introduce the normalized functions
\begin{equation}
 \varphi_i(\mathbf u)=\frac1{\sqrt\pi}
 (1+|\mathbf u-\mathbf y_i|^2)^{-1+iE},\qquad
 \fnNorm{\varphi_i}_{L^2(\mathbb R^2)}=1.
\end{equation}
Then
\begin{equation}
 \boxed{F_{ij}=\int_{\mathbb R^2}
 \overline{\varphi_i(\mathbf u)}\varphi_j(\mathbf u)\,d^2\mathbf u.}
 \label{fn:dyn:eq:gram}
\end{equation}
Consequently $z^\dagger Fz=\fnNorm{\sum_i z_i\varphi_i}_{L^2}^2\geq0$ for
every finite vector $z$ and every $d>0$.

To evaluate the overlap, use translation and rotation to put the two centers
at separation $r=|\mathbf y_i-\mathbf y_j|$. With $a=1+iE$, $b=1-iE$, the
beta-parameter identity followed by the planar integral gives
\begin{align}
 \frac1\pi\int_{\mathbb R^2}
 \frac{d^2\mathbf u}{(1+|\mathbf u|^2)^a(1+|\mathbf u-\mathbf r|^2)^b}
 &=\frac1{\Gamma(a)\Gamma(b)}
 \int_0^1\frac{t^{iE}(1-t)^{-iE}}{1+r^2t(1-t)}\,dt\\
 &=\frac1{2|\Gamma(1+iE)|^2}
 \int_{-\infty}^{\infty}\frac{e^{iEv}}{\cosh v+\cosh\chi}\,dv,
 \qquad \chi=2\operatorname{arsinh}(r/2).
\end{align}
The first identity follows from $a+b=2$ and
$\int_{\mathbb R^2}(|\mathbf u|^2+C)^{-2}d^2\mathbf u=\pi/C$;
the second uses $v=\log[t/(1-t)]$ and $\cosh\chi=1+r^2/2$.
All the parameter integrals converge absolutely before evaluating the Fourier
integral. For $\chi>0$, a rectangular contour of height $2\pi i$, with poles
at $v=\pm\chi+i\pi$, yields
\begin{equation}
 \int_{-\infty}^{\infty}\frac{e^{iEv}}{\cosh v+\cosh\chi}\,dv
 =\frac{2\pi\sin(E\chi)}{\sinh(\pi E)\sinh\chi}.
\end{equation}
Indeed, the sum of residues is
$-2i e^{-\pi E}\sin(E\chi)/\sinh\chi$, and the upper edge multiplies the
real-axis integral by $e^{-2\pi E}$. Using
$|\Gamma(1+iE)|^2=\pi E/\sinh(\pi E)$ proves
Eq.~\eqref{fn:dyn:eq:gram}. The coincident-point value follows by continuity and is one.

\section*{Methodological note on AI assistance}
\phantomsection
\addcontentsline{toc}{section}{Methodological note on AI assistance}
\label{sec:AIassistance}

The authors used OpenAI ChatGPT 6.1 Sol  to assist with analytical
derivations and consistency checks, drafting and revising
scientific explanations, developing and debugging numerical
code, and performing numerical cross-checks. The tool also
assisted with revising the plotting code and preparing
Fig.~\ref{fig:largeNparity}. The authors specified the physical
model, assumptions, parameter ranges, and quantities to be
evaluated, and directed the subsequent revisions. They assessed
the assisted contributions by reviewing the analytical steps
and their assumptions, inspecting the numerical implementations,
checking consistency between the reported results and the
computational outputs, and comparing the figures with the
underlying numerical data. AI-generated checks were treated
as supporting calculations rather than independent validation.
The authors reviewed and edited the final text and retain full
responsibility for the methods, results, interpretations,
and presentation.

\addcontentsline{toc}{section}{References}
\bibliographystyle{apsrev4-2}
\makeatletter\immediate\write\@auxout{\string\citation{apsrev42Control}}\makeatother
\IfFileExists{SM.bib}{\bibliography{SM}}{\bibliography{manuscript/SM}}